\documentclass[amsmath,amssymb,longbibliography,floatfix,showpacs,notitlepage,nofootinbib,superscriptaddress,twocolumn]{revtex4-1}

\usepackage[T1]{fontenc}
\usepackage[utf8]{inputenc}
\usepackage{bm}
\usepackage{hyperref}
\usepackage{graphicx}
\graphicspath{{figures/}}
\usepackage{color}
\hypersetup{colorlinks,
            citecolor=black,
            filecolor=black,
            linkcolor=black,
            urlcolor=blue }

\newcommand{\pol}{\mathbf{P}}
\newcommand{\polb}{\bar{\mathbf{P}}}

\newcommand{\hv}{\protect{\hat{v}}}

\newcommand{\thetav}{\theta_\text{v}}

\newcommand{\dd}{\protect{\text{d}}}
\newcommand{\ri}{\protect{\text{i}}}
\newcommand{\rv}{\protect{\vec{r}}}
\newcommand{\pv}{\protect{\vec{p}}}
\newcommand{\Bv}{\mathbf{B}}

\newcommand{\Vv}{\mathbf{V}}

\newcommand{\be}{\mathbf{e}}
\newcommand{\GF}{G_{\text{F}}}
\newcommand{\rini}{r_\text{ini}}
\newcommand{\rfin}{r_\text{fin}}

\newcommand*{\UNM}{Department of Physics \& Astronomy, University of New
  Mexico, Albuquerque, NM 87131, USA}

\newcommand*{\LANL}{Theoretical Division, Los Alamos National Laboratory, %
Los Alamos, NM 87545, USA}

\begin{document}

\title{Spectral swaps in a two-dimensional neutrino ring model}

\author{Joshua D.\ Martin}
\email{josh86@unm.edu}
\affiliation{\UNM}

\author{J.\ Carlson}
\affiliation{{\LANL}}

\author{Huaiyu Duan}
\affiliation{\UNM}

\date{\today}

\begin{abstract}
  Neutrinos emitted deep within a supernova explosion experience a self-induced index of refraction.  In the stationary, one-dimensional (1D) supernova ``bulb model'', this self-induced refraction can lead to a collective flavor transformation which is coherent among different neutrino momentum modes. Such collective oscillations can produce partial swaps of the energy spectra of different neutrino species as the neutrinos stream away from the proto-neutron star. However, it has been demonstrated that the spatial symmetries (such as the spherical symmetry in the bulb model) can be broken spontaneously by collective neutrino oscillations in multi-dimensional models. Using a stationary, 2D neutrino ring model we demonstrate that there exist two limiting scenarios where collective oscillations may occur. In one limit, the collective flavor transformation begins at a radius with relatively high neutrino densities and develops small-scale flavor structures. The loss of the spatial correlation in the neutrino flavor field results in similar (average) energy spectra for the anti-neutrinos of almost all energies and the neutrinos of relatively high energies. In the other limit, the flavor transformation starts at a radius where the neutrino densities are smaller (e.g., due to the suppression of the high matter density near the proto-neutron star). Although the spatial symmetry is broken initially, it is restored as the neutrino densities decrease, and the neutrinos of different flavors partially swap their energy spectra as in the 1D bulb model. This finding may have interesting ramifications in other aspects of supernova physics.
\end{abstract}

\maketitle

\section{Introduction} \label{sec:intro}

Neutrinos are fundamental particles which are nearly massless, carry no electric charge,
and interact only through the weak interaction and gravitation.
Nevertheless, they can have a significant impact on  the evolution of their environments such as the early universe, core-collapse
supernovae and neutron star mergers where they are copiously produced.

Neutrinos are emitted in weak-interaction states which are not
coincident to the mass states of the free particle.  As a result,
neutrinos can undergo flavor oscillations even in vacuum \cite{Tanabashi:2018oca}.
The flavor evolution of a neutrino becomes more complicated when it propagates
through a dense medium.  In the coherent forward
scattering limit, the neutrino acquires an index of refraction which depends on the
leptonic flavor content of the medium
\cite{Mikheev:1986gs, Wolfenstein:1977ue, Wolfenstein:1979ni}.  This effect becomes even more interesting when there is a
significant presence of other neutrinos, as the coherent
forward scattering permits the propagating neutrinos to exchange flavor information with background neutrinos
\cite{1987ApJ...322..795F,Notzold:1987ik,Pantaleone:1992xh}.

The flavor evolution of a dense neutrino medium is governed by a system of seven dimensional (1 in time, 3 in space, and 3 in neutrino momentum), nonlinear, partial-differential equations that have been tackled with various simplifications.
The well
known ``bipolar model'' assumes a homogeneous and isotropic distribution of mono-energetic neutrinos \cite{Kostelecky:1994dt,Duan:2005cp}.  This
model exhibits a behavior which is isomorphic to a gyroscopic pendulum \cite{Hannestad:2006nj}.  The ``bulb model'' is a spherically symmetric and
stationary model which simulates neutrinos streaming off a single spherical surface
\cite{Duan:2006an}. These models have been studied extensively in the literature, and
display a wealth of interesting features such as spectral swaps/splits, and coherence
across momentum modes over long distance scales (see, e.g., Refs.~\cite{Duan:2006jv, Fogli:2007bk, Duan:2007bt, EstebanPretel:2007ec, Dasgupta:2009mg, Friedland:2010sc} among many others;
see also Ref.~\cite{Duan:2010bg} for a review).

The results of the simplified models, however, are not necessarily physical because of the symmetric conditions that are artificially imposed to simplify the calculations. Indeed,
it has been shown through the stability analysis of the linearized flavor evolution equations that these symmetries can be spontaneously broken by neutrino oscillations \cite{Raffelt:2013rqa, Duan:2014gfa, Chakraborty:2015tfa, Abbar:2015fwa}. The few existing numerical calculations confirm that the symmetry-breaking instabilities can indeed trigger rapid oscillations of the neutrino flavor field on small distance scales and thus reduce or even destroy its correlation in time, space and momentum
\cite{Mirizzi:2013rla, Mangano:2014zda, Mirizzi:2015fva, Mirizzi:2015hwa, Capozzi:2016oyk, Martin:2019kgi}.
(However, see Ref.~\cite{Martin:2019gxb} where a coherent fast oscillation wave was reported for a dynamic model.)

The primary goal of this work is to investigate if the spectral swap, the hallmark of collective neutrino oscillations in the one (spatial) dimensional (1D) model, also exists in multi-dimensional models. For this purpose we adopt the a stationary 2D neutrino ring model similar to that in Ref.~\cite{Mirizzi:2015hwa} except for trading the multiple angle bins for multiple energy bins.

The rest of the paper is organized as follows.
In Sec.~\ref{sec:ringmodel}, we write down the equations of
motion that govern the neutrino oscillations in the ring model.  In Sec.~\ref{sec:approach} we describe the configurations of the numerical examples and the approach that we used to solve them. In Sec.~\ref{sec:results} we
present the results of our calculations, and in Sec.~\ref{sec:conclusions}, we discuss the implications of our results and give our conclusions.

\section{Equations of motion} \label{sec:ringmodel}

We consider the mixing of two neutrino flavors, $\nu_{e}$
and $\nu_{\tau}$, where $\nu_\tau$ is a suitable linear combination of the physical $\nu_\mu$ and $\nu_\tau$.  In the absence of collisions, the flavor contents of the neutrino and antineutrino fields of momentum $\pv$ at spacetime point $(t,\rv)$ can be represented by the normalized
polarization vectors $\pol_{\pv}(t,\rv)$ and $\polb_{\pv}(t,\rv)$ in flavor space, respectively.
The evolution of the polarization vectors is governed by the system of transport equations
\cite{Sigl:1992fn}
\begin{subequations}
    \label{eq:eom-full}
  \begin{align}
    (\partial_t + \hv \cdot \overrightarrow \nabla) \pol_{\pv}
      &= (\omega\Bv + \lambda \be_3 + \Vv_{\hv})\times \pol_{\pv},\\
    (\partial_t + \hv\cdot\overrightarrow\nabla) \polb_{\pv}
      &= (-\omega\Bv + \lambda \be_3 + \Vv_{\hv})\times \polb_{\pv},
  \end{align}
\end{subequations}
where the three terms inside the parentheses on the righthand side of the equation represent the vacuum mixing, the matter effect, and the neutrino self-coupling, respectively. The vacuum oscillation frequency and the matter potential are $\omega = \delta m^{2} / 2 E$ and $\lambda = \sqrt{2} \GF n_{e}$, respectively, where $\delta m^2$ and $E=|\pv|$ are the mass-squared difference and the energy of the neutrino, $\GF$ is the Fermi coupling constant,
and $n_{e}$ is the local net number density of the electrons. The flavor basis vectors $\be_i$ ($i=1,2,3$) are defined in such a way that the vacuum mixing is given by the unit vector
$\Bv = \sin(2 \thetav) \be_{1}  - \cos(2 \thetav) \be_{3}$ with $\thetav$ being the vacuum mixing angle. In this basis,
\begin{align}
      \mathcal{P}_{\nu_e\nu_e} = \frac{1+P_3}{2} = \frac{1+\pol\cdot\be_3}{2}
\end{align}
is the probability for a neutrino initially in the electron flavor to survive in the same flavor or the $\nu_e$ survival probability.
Lastly, the neutrino self-interaction or neutrino-neutrino forward scattering is given by
\begin{widetext}
  \begin{align}
    \Vv_\hv =
          \sqrt2\GF \int\!\frac{\dd^3 p'}{(2\pi)^3}
        (1-\hv\cdot\hv')\{[\rho^0_{ee}(\pv') - \rho^0_{\tau\tau}(\pv')]\pol_{\pv'}
          -[\bar\rho^0_{ee}(\pv') - \bar\rho^0_{\tau\tau}(\pv')]\polb_{\pv'}\},
  \end{align}
  \end{widetext}
where $\rho^0_{\beta\beta}$ and $\bar\rho^0_{\beta\beta}$ ($\beta=e, \tau$) are the initial occupation numbers of the corresponding neutrino flavor. We use the superscript 0 to denote the values of the physical quantities when there were no neutrino flavor transformation.

We utilize a stationary, 2D neutrino ring model
 similar to those in Refs.~\cite{Shalgar, Mirizzi:2015hwa}. In this model, the neutrinos are constantly emitted from a ring of radius $R_\nu$ and stream freely in the $x$-$y$ plane.
In the current work, we assume only two neutrino beams are emitted from every point on the neutrino ring such that
\begin{align}
  \rho^0_{ee}(\pv) = \frac{n^0_{\nu_e}}{2E} f^0_{\nu_e}(E)
\delta(p_z) [\delta(\vartheta - \vartheta_0) + \delta(\vartheta + \vartheta_0)]
\end{align}
on the neutrino ring, where $n^0_{\nu_e}$ and $f^0_{\nu_e}(E)$ are the number density and the normalized energy distribution of $\nu_e$ on the ring, respectively,
 $\vartheta\in[-\pi/2,\pi/2]$ is the angle that the neutrino momentum $\pv$ makes with the radial direction, and $\vartheta_0>0$ is a constant. We assume similar  emissivities for the other neutrino flavors with the following normalization conditions:
 \begin{align}
 \int_0^\infty f_\nu^0(E)\,\dd E=\frac{n_\nu^0}{n_{\nu_e}^0}  \qquad (\nu=\nu_e,\bar\nu_e,\nu_\tau,\bar\nu_\tau).
 \label{eq:f-norm}
 \end{align}

In this stationary, two-beam neutrino ring model,
\begin{align}
  \pol_{\pv}(t,\rv) \longrightarrow \pol^\pm_E(r, \Phi),
  \quad \polb_{\pv}(t,\rv) \longrightarrow \polb^\pm_E(r, \Phi),
\end{align}
and
\begin{align}
  \partial_t + \hv\cdot\vec\nabla \longrightarrow D_\pm=
  v_r \partial_r \pm \frac{R_\nu}{r^2}\sin\vartheta_0 \partial_\Phi,
\end{align}
where we have used the polar coordinates $(r, \Phi)$ in the $x$-$y$ plane with the center of the neutrino ring at the origin, the plus and minus signs are for the neutrino beams with emission angles $\pm\vartheta_0$, respectively, and
\begin{align}
  v_r (r) = \sqrt{1 - \left(\frac{R_\nu}{r}\right)^2\sin^2\vartheta_0}
\end{align}
is the radial component of the neutrino velocity.
The neutrino self-coupling potential becomes
\begin{align}
    \Vv_\hv \rightarrow\Vv_\pm(r,\Phi)
    &= \mu \int_0^\infty
    [(f_{\nu_e}^0- f_{\nu_\tau}^0)\pol_E^\mp
\nonumber\\
    &\quad- (f_{\bar\nu_e}^0-f_{\bar\nu_\tau}^0) \polb_E^\mp]\,\dd E,
\end{align}
where
\begin{align}
  \mu(r) = \sqrt2 \GF n^0_{\nu_e} \frac{\sin^2\vartheta_0\cos\vartheta_0}{\sqrt{1- (R_\nu/r)^2\sin^2\vartheta_0}}\left(\frac{R_\nu}{r}\right)^4.
  \label{eq:mu}
\end{align}
In Eq.~\eqref{eq:mu} we have modified the ring model by including an extra factor of $R_\nu/r$ to mimic 3D supernova models in which the neutrino fluxes decrease as $1/r^2$ instead of $1/r$ in the original ring model.
We also assume a large matter density distributed symmetrically about the center the neutrino ring such that
\begin{align}
  \pm\omega\Bv + \lambda \be_3 \longrightarrow \mp\eta |\omega|\cos2\thetav\be_3
  \label{eq:matt-eff}
\end{align}
in the appropriate rotating reference frame in flavor space, where $\eta=+1$ for the normal neutrino mass hierarchy (NH) and $-1$ for the inverted hierarchy (IH).
In summary, the original equations of motion \eqref{eq:eom-full} reduce to
\begin{subequations}
    \label{eq:eom}
  \begin{align}
    D_\pm \pol^\pm_E
      &= ( -\eta |\omega|\cos2\thetav\be_3 + \Vv_\pm)\times \pol^\pm_E,\\
      D_\pm \polb^\pm_E
        &= ( \eta |\omega|\cos2\thetav\be_3 + \Vv_\pm)\times \polb^\pm_E
  \end{align}
\end{subequations}
for the model that we are considering. It is straightforward to show from the above the equations that the average electron lepton number (ELN),
\begin{widetext}
\begin{align}
  \mathcal{L} = \int_0^{2\pi} \frac{\dd \Phi}{2\pi}
  \int_0^\infty \dd E\, [(P_{E,3}^+ + P_{E,3}^-)(f_{\nu_e}^0 - f_{\nu_\tau}^0) -
  (\bar{P}_{E,3}^+ + \bar{P}_{E,3}^-)(f_{\bar\nu_e}^0 - f_{\bar\nu_\tau}^0)],
  \label{eq:L}
\end{align}
\end{widetext}
is constant along $r$.

\section{Numerical approach} \label{sec:approach}

As a concrete example, we consider a model with a neutrino emission ring of radius $R_{\nu} = 10\, \text{km}$. Two neutrino beams with emission angles  $\pm\pi/4$ are emitted from each point on the ring with an approximate circular symmetry around the ring. As in Ref.~\cite{Duan:2006an}, we assume the neutrino species $\nu$ in each neutrino beam has the Fermi-Dirac spectrum
\begin{equation}
  f_{\nu}^0(E) \propto \frac{ E^{2}}{\exp(E/T_\nu - \xi_{\nu}) + 1}
\end{equation}
with $T_{\nu_{e}} = 2.76 \, \text{MeV}$, $T_{\bar{\nu}_{e}} = 4.01 \, \text{MeV}$,
$T_{\nu_\tau} = T_{\bar{\nu}_\tau}= 6.26 \, \text{MeV}$,
and $\xi_\nu=3$ for all the neutrino species.
We also assume
 $ \mu(R_\nu) = 5\times10^{4} \, \text{km}^{-1}$, $n_{\bar\nu_e}^0/n_{\nu_e}^0 = 0.8$ and $n_{\nu_\tau}^0/n_{\nu_e}^0=n_{\bar\nu_\tau}^0/n_{\nu_e}^0  = 0.4$, and
$|\delta m^{2}|\cos(2\thetav) = 2.5 \times 10^{-15} \, \text{MeV}^{2}$
for vacuum mixing.

\begin{figure}[tbh]
  \begin{center}
    \includegraphics*[width=\columnwidth]{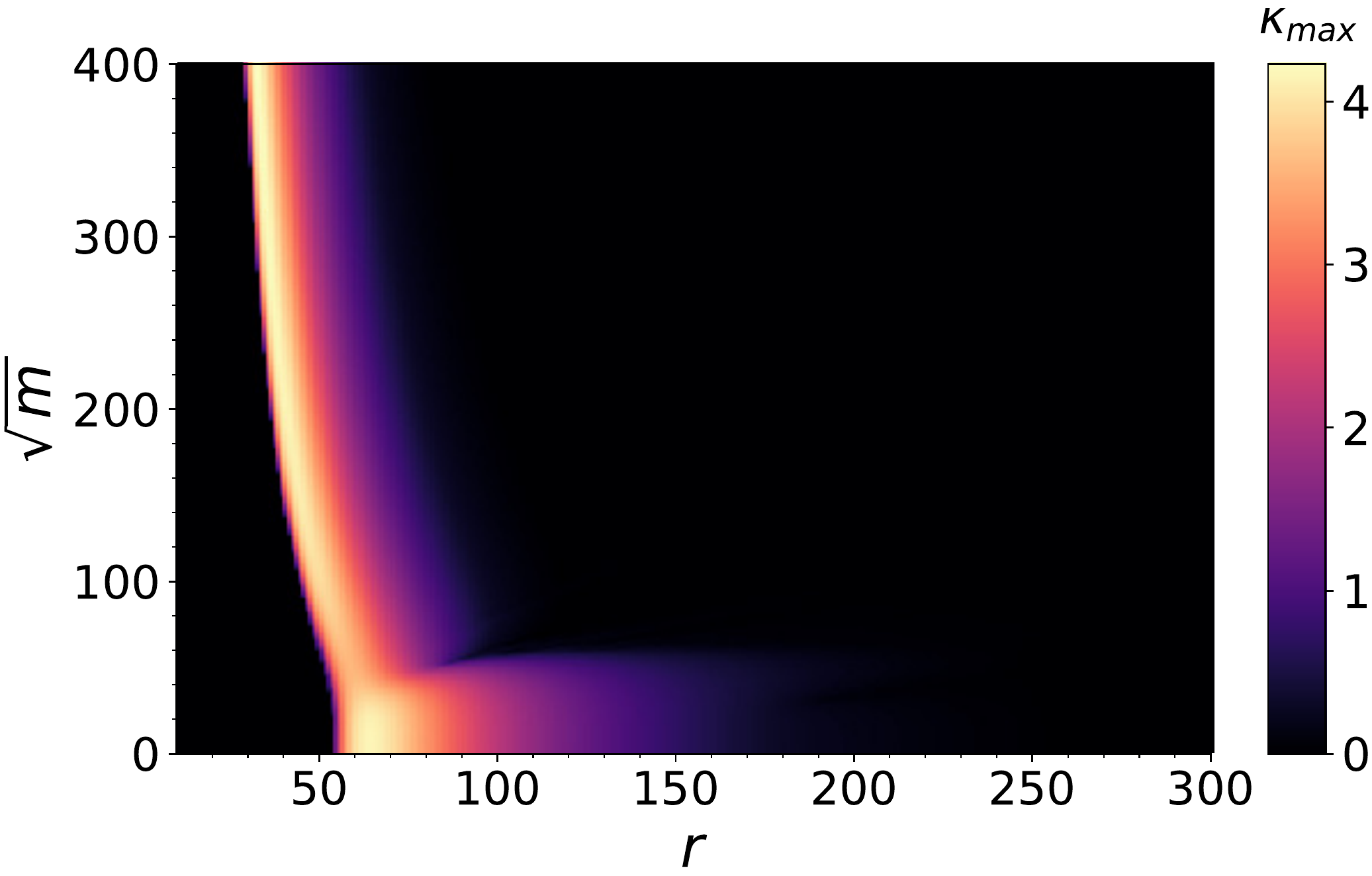}
  \end{center}
  \caption{(Color online)
  The maximum exponential growth rate $\kappa_\text{max}$ as a function of the Fourier moment index $m$ and the radius $r$ in the two-beam neutrino ring model. This growth rate is independent of the neutrino mass hierarchy for this model.
  }\label{fig:kappa}
\end{figure}

One way to solve Eq.~\eqref{eq:eom} is to first perform the Fourier transformation:
\begin{subequations}
\begin{align}
  \pol^{(m)}(r) &= \int_0^{2\pi} \pol(r,\Phi) e^{-\ri m \Phi}
  \frac{\dd\Phi}{2\pi}, \\
  \polb^{(m)}(r) &= \int_0^{2\pi} \polb(r,\Phi) e^{-\ri m\Phi}
  \frac{\dd\Phi}{2\pi},
\end{align}
\end{subequations}
and then solve the evolution of the Fourier moments numerically \cite{Mirizzi:2015hwa}.
In the linear regime where $|\pol-\be_3|$ and $|\polb-\be_3|$ are much less than 1,  one can perform the  flavor stability analysis on the neutrino gas \cite{Chakraborty:2015tfa}. In this regime, the evolution of the Fourier moments of different values of $m$ are decoupled. In Fig.~\ref{fig:kappa} we show the maximum exponential growth rate $\kappa_\text{max}$ for various Fourier moments and at different radii assuming that no significant flavor conversion has occurred.

In Ref.~\cite{Mirizzi:2015hwa} all the Fourier moments but those with $m=0$ and $\pm1$ were assumed to be 0 initially. Fig.~\ref{fig:kappa} shows that a calculation with this assumption would not see flavor conversion until $r\gtrsim 50$ km where the lowest moments become unstable, even though the higher moments are unstable at smaller radii. It has been shown in the 2D neutrino line model, which is similar to the ring model but has constant neutrino densities, that the high moments that are unstable can not only grow by themselves but also cause the growth of other Fourier moments, and eventually lead to significant flavor conversion in the whole system \cite{Martin:2019kgi}.

In this work we solve Eq.~\eqref{eq:eom} directly in the polar coordinates $(r, \Phi)$. We discretize both $\Phi\in[0,2\pi)$ and $E\in[0, 75]$ MeV into pre-determined, equal-sized, discrete bins, and we solve the corresponding polarization vectors adaptively along the radial direction using a finite difference algorithm derived from the Lax-Wendroff
method \cite{NR2002, Martin:2019kgi}.
We use a large number of $\Phi$ bins (128,000 for most of the calculations) to ensure the numerical convergence and the accuracy of the results. We use a relatively small number (128) of energy bins  because collective neutrino oscillations are known to be insensitive to the energy resolution. We do not enforce the unitary condition $|\pol| = |\polb| = 1$ but use it as a validity check of the calculations.

Although the neutrino gas has flavor instabilities even on the neutrino ring in the two-beam model, this is not necessarily the case for the 2D and 3D models with continuous angular distributions in neutrino emission. It has been shown that these instabilities can be suppressed near the proto-neutron star by the presence of a large matter density \cite{Chakraborty:2015tfa}.
(The largest index of the unstable moments is underestimated in Ref.~\cite{Chakraborty:2015tfa} because of the absence of a factor of $\beta_\text{max}^2$ in its Eq.~(2.6).) To mimic this suppression in the supernova environment, we start the calculations at various radii $\rini$ in the two-beam ring model.
In this work we study six cases with different neutrino mass hierarchies and various values of $\rini$ which are listed in Table~\ref{tab:cases}. We stop the calculations at $\rfin=350$ km where the neutrino fluxes are too small to affect oscillations.
\begin{table}[htb]
\caption{\label{tab:cases} The parameters of the six cases in the numerical survey of the two-beam neutrino ring model.}
\begin{ruledtabular}
  \begin{tabular}{cccc}
    Case No. & Hierarchy & $\rini$ (km) & $\Phi$ bins \\ \hline
    I & IH & 105 & 128,000 \\
    II & IH & 120 & 128,000  \\
    III & IH & 140 & 128,000  \\
    IV & NH & 120 & 256,000 \\
    V & NH & 130 & 128,000  \\
    VI & NH & 140 & 128,000
  \end{tabular}
\end{ruledtabular}
\end{table}

In all six cases, we assume the following energy-independent initial conditions for the polarization vectors:
\begin{equation} \label{eq:polInit}
  \pol^{\pm}_E(\rini,\Phi) = \polb^{\pm}_E(\rini,\Phi) =
    \left[ \epsilon_{\pm}, 0, \sqrt{1 - \epsilon_{\pm}^{2} } \right],
\end{equation}
where
\begin{align}
  2\epsilon_-(\Phi) = \epsilon_+(\Phi) = \epsilon_0 + \epsilon_1 \sin(\Phi)
  + \epsilon_\text{g} e^{-(\Phi-\pi)^2/2\sigma^2}
  \label{eq:pert}
\end{align}
with $\epsilon_0 = 10^{-3}$, $\epsilon_1 = 10^{-4}$, $\epsilon_\text{g} = 10^{-3}$,
and $\sigma^{2} = 0.1$. We have intentionally broken the symmetry between the two neutrino beams emitted from the same point so that the neutrino gas can be unstable in both mass hierarchies even for the $m=0$ mode. We have also included perturbations which break the circular symmetry both globally and locally.

\section{Flavor evolution}\label{sec:results}

\subsection{Inverted mass hierarchy}
\newcommand{\figscalei}{0.47}
\begin{figure*}[htb]
  \begin{center}
    $\begin{array}{@{}l@{\hspace{0.01in}}l@{\hspace{0.01in}}l@{}}
      \includegraphics*[scale=\figscalei]{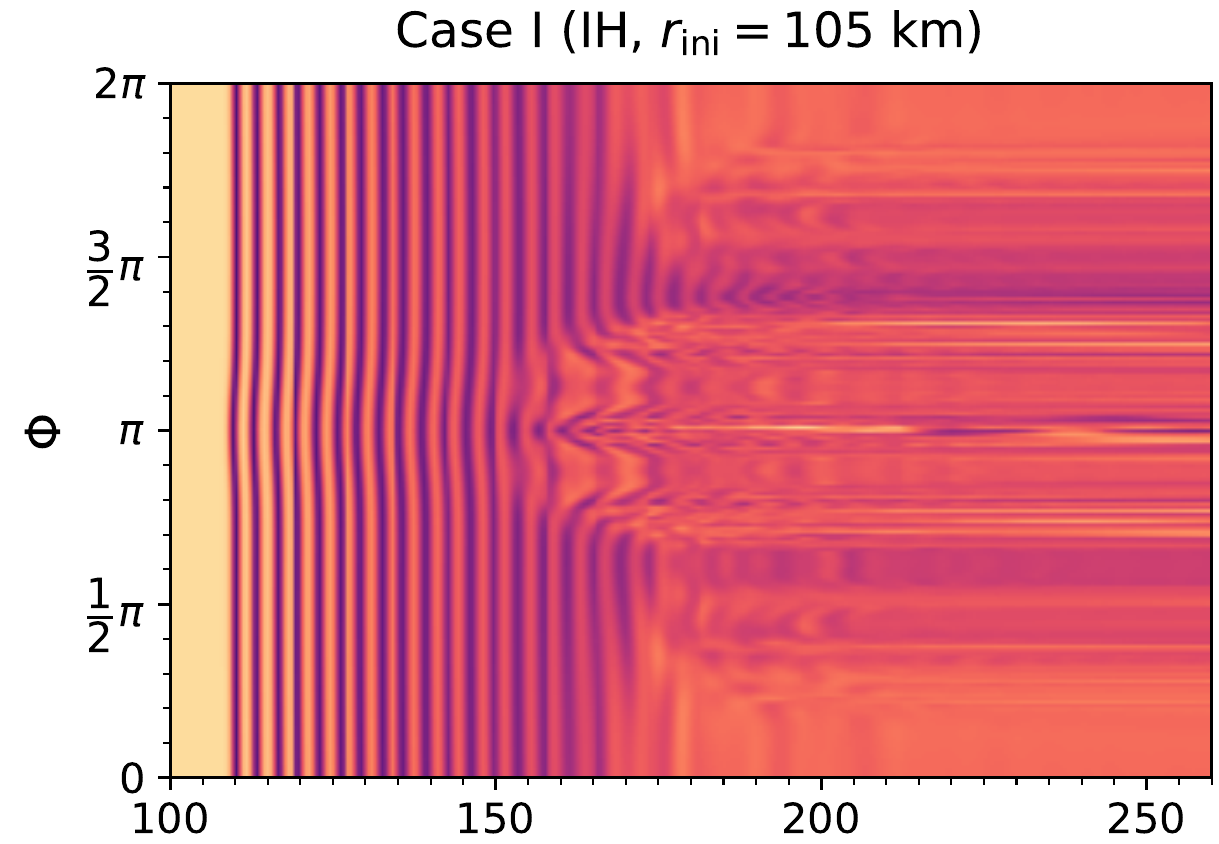} &
      \includegraphics*[scale=\figscalei]{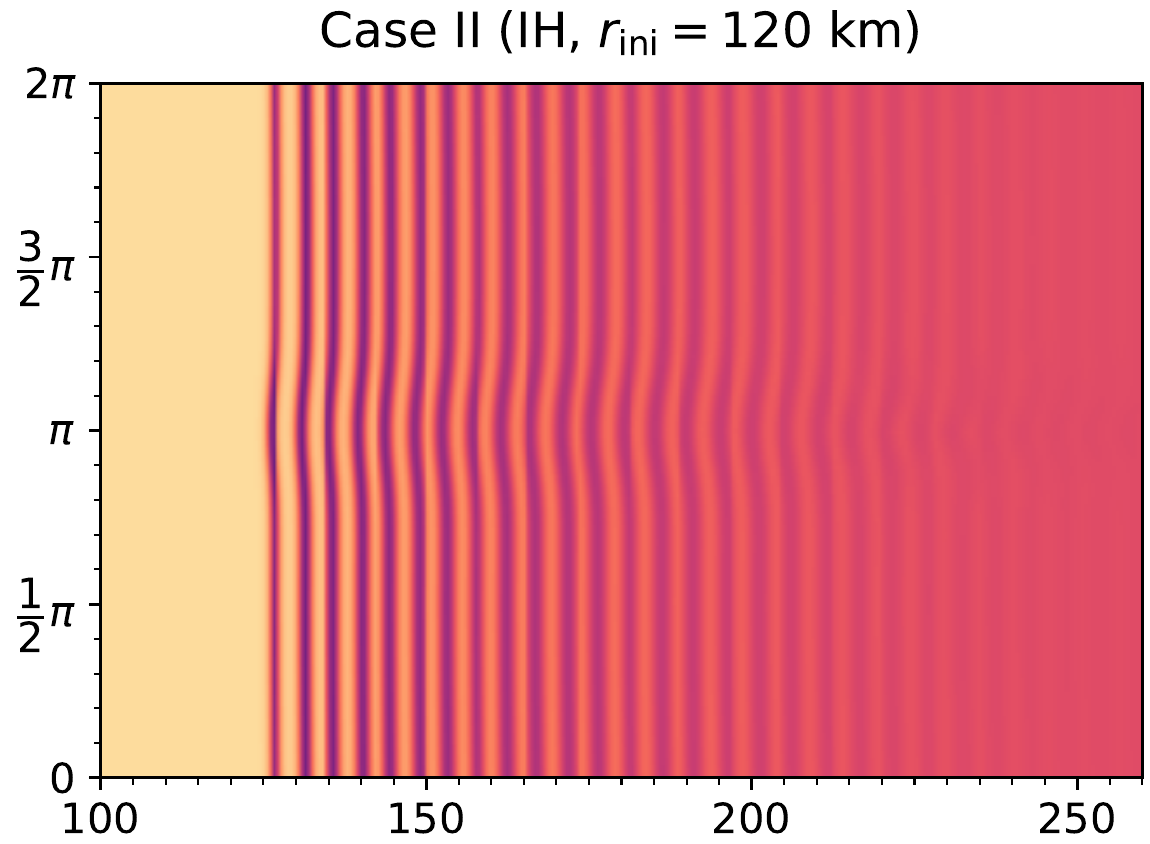} &
      \includegraphics*[scale=\figscalei]{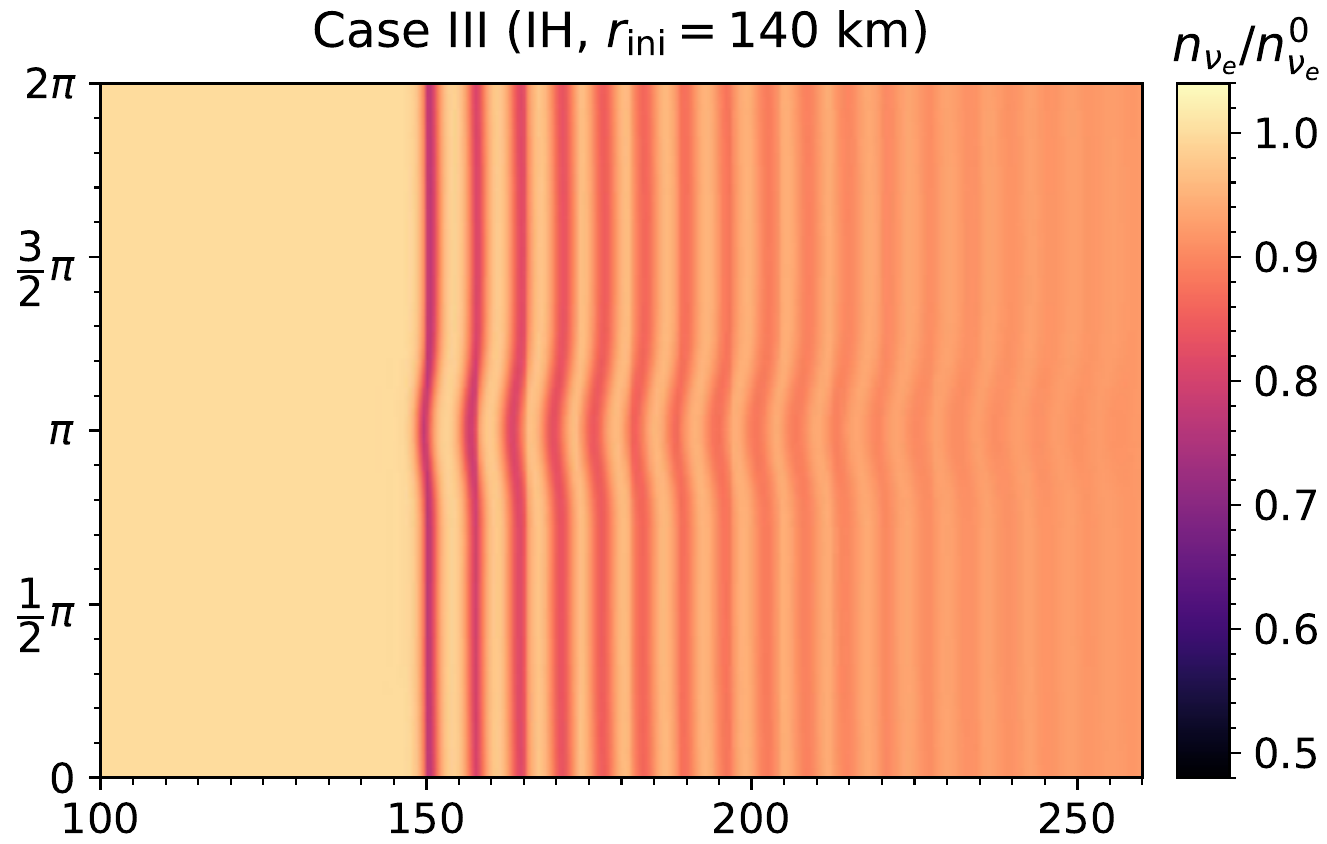} \\
      \includegraphics*[scale=\figscalei]{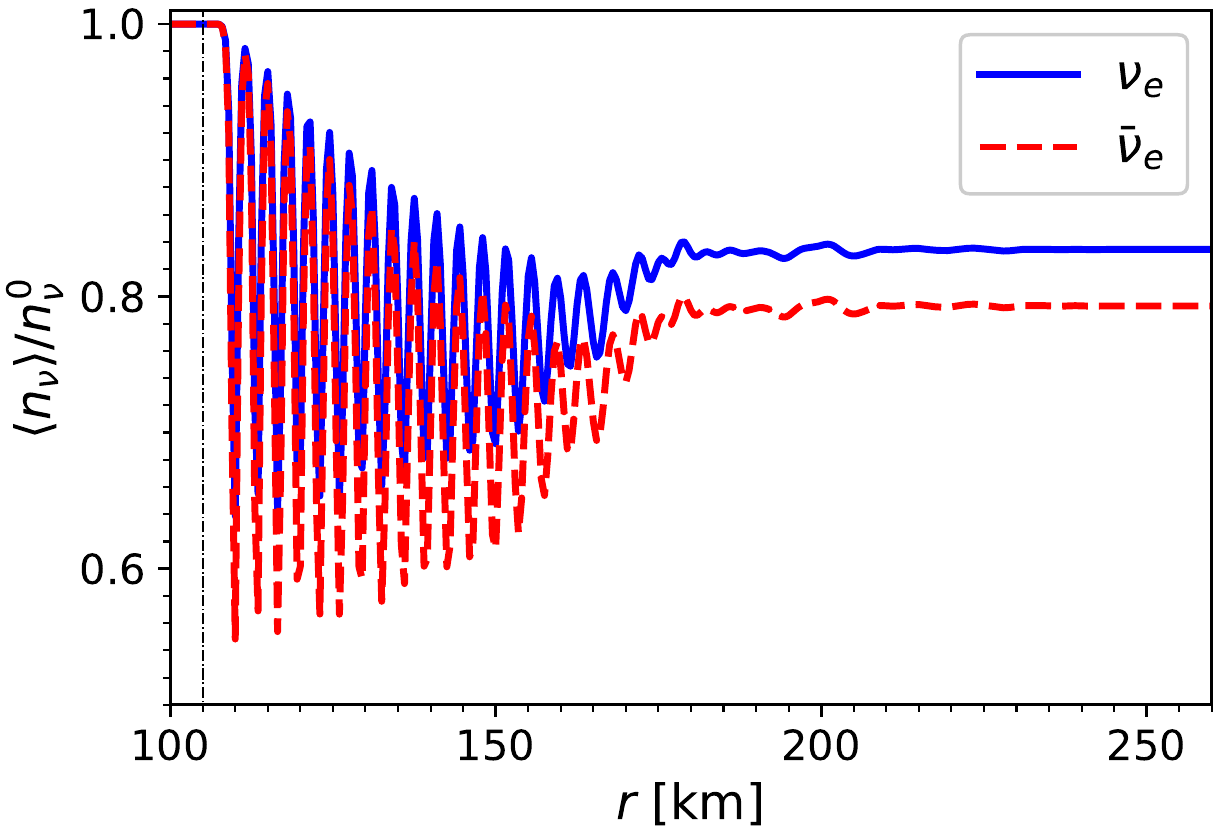} &
      \includegraphics*[scale=\figscalei]{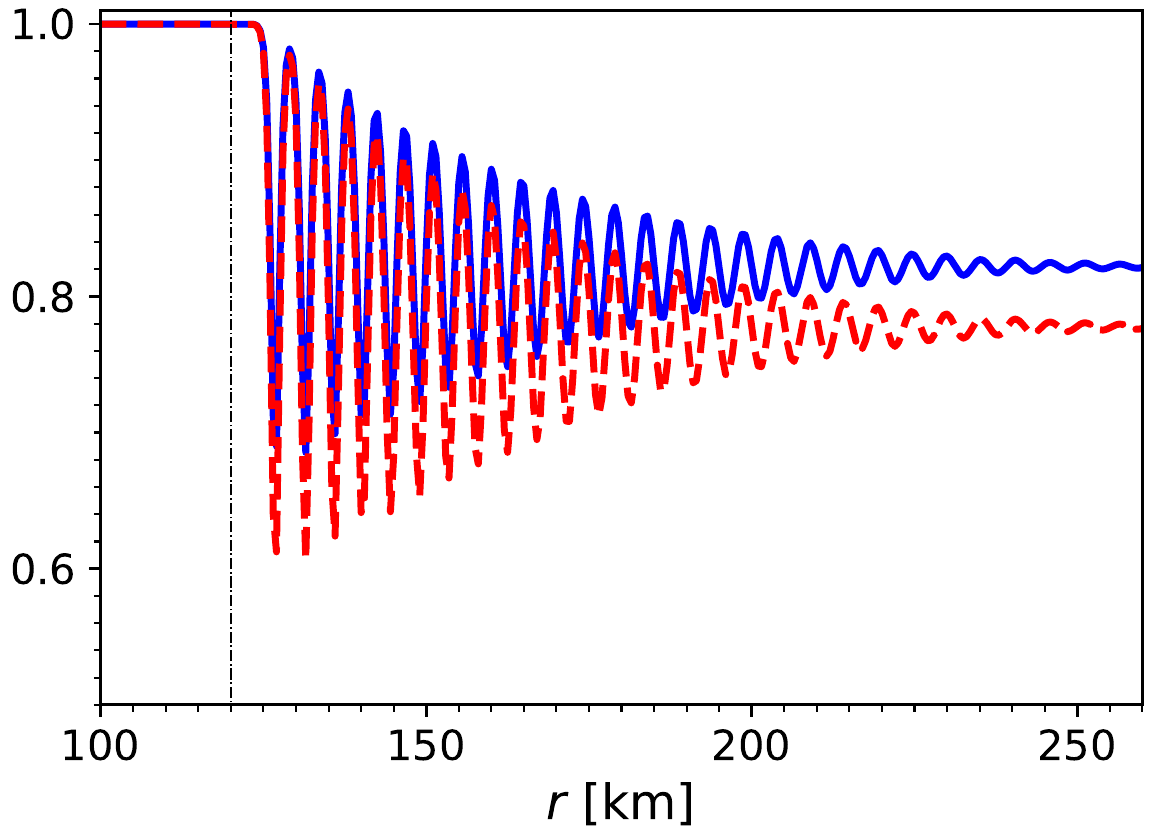} &
      \includegraphics*[scale=\figscalei]{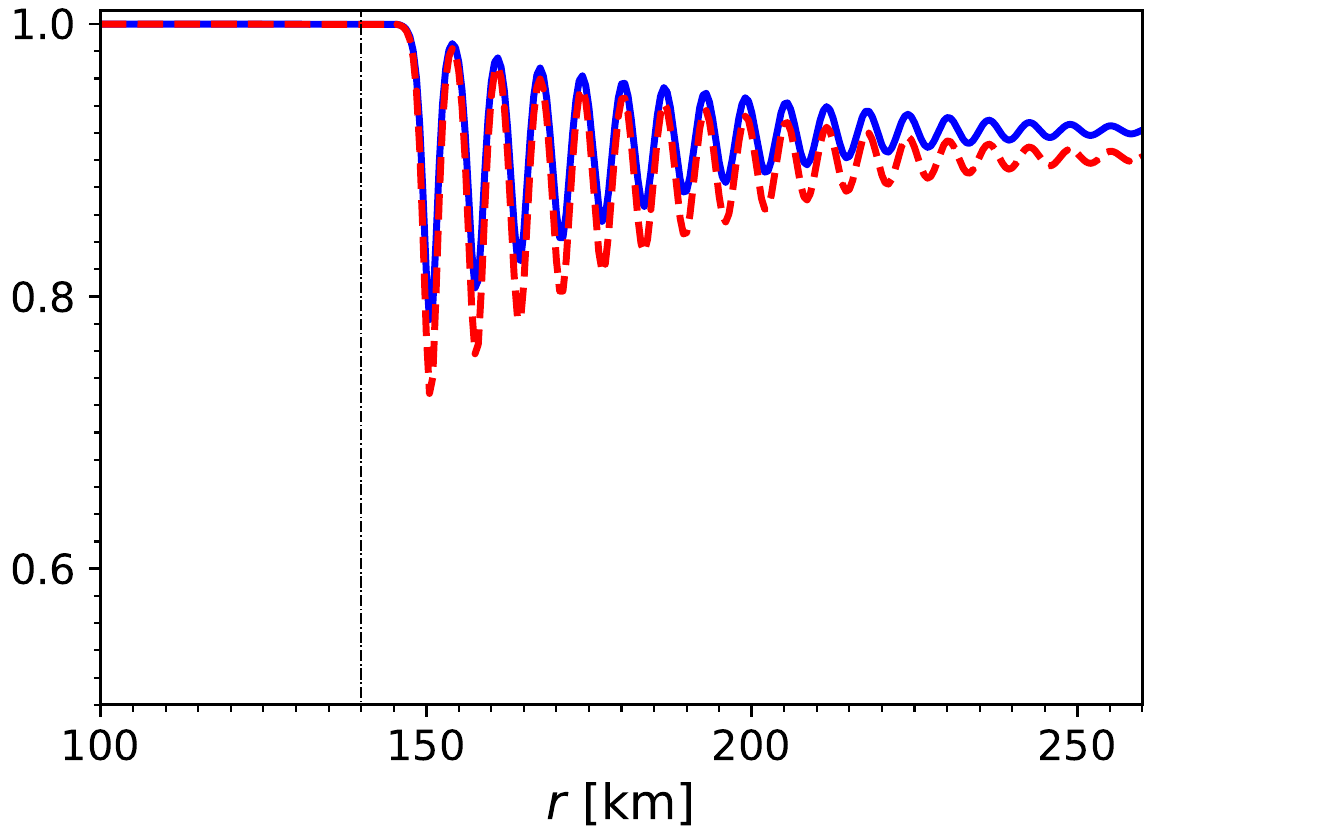}
    \end{array}$
  \end{center}
  \caption{(Color online)
  Top panels: The ratios of local $\nu_e$ densities with and without flavor conversions, $n_{\nu_e}/n_{\nu_e}^0$, in the polar coordinates $(r,\Phi)$ in cases I through III which all employ the inverted neutrino mass hierarchy (IH) and have $\rini=105$ km (left panel), 120 km (middle panel), and 140 km (right panel), respectively. Bottom panels: The ratios $\langle n_{\nu_e}^0\rangle/n_{\nu_e}^0$ and $\langle n_{\bar\nu_e}^0\rangle/n_{\bar\nu_e}^0$  averaged over $\Phi$ at the same $r$. The values of $\rini$, the radii where the calculations begin,
  are shown as the vertical dot-dashed lines in the bottom panels.
  }\label{fig:nnu-IH}
\end{figure*}

In the upper panels of Fig.~\ref{fig:nnu-IH} we show the ratios of local $\nu_e$ densities with and without flavor conversions,
\begin{align}
   \frac{n_{\nu_e}}{n_{\nu_e}^0}
&= \frac{1}{4}\int_0^\infty [ (P_{E,3}^+ + P_{E,3}^-) (f^0_{\nu_e} - f^0_{\nu_\tau})
\nonumber \\
&\quad + 2 (f^0_{\nu_e} + f^0_{\nu_\tau})]\,\dd E,
\end{align}
in cases I through III with the IH which mimic the scenarios where the matter suppression is lifted at radii $\rini=105$ km, 120 km, and 140 km, respectively. We also show the ratios $\langle n_{\nu_e} \rangle / n_{\nu_e}^0$ and $\langle n_{\bar\nu_e} \rangle / n_{\bar\nu_e}^0$ averaged over the angular coordinate $\Phi$ in the lower panels of the same figure.
Similar to the 1D bulb model \cite{Duan:2006jv,Duan:2006an}, the neutrino gases in the 2D ring model exhibit bipolar-like oscillations in all three cases shortly after the calculations begin. The oscillation length scales increase with $r$ because the bipolar oscillation have frequency
$\propto \sqrt{\mu(r)}$
\cite{Kostelecky:1994dt,Duan:2005cp} which decreases with $r$ in the ring model. These oscillations at different $\Phi$ are coherent initially, although the circular symmetry is broken as manifested by the slight $\Phi$ dependence of $n_{\nu_e}/ n_{\nu_e}^0$ in the upper panels of the figure.  Similar to the neutrino gases in the line model \cite{Martin:2019kgi},
small flavor structures begin to develop at $r\approx 160$ km in case I, and $\langle n_{\nu} \rangle / n_{\nu}^0$
reach equilibrium values at $r\approx 180$ km after the small-scale flavor structures are sufficiently developed. In contrast, the coherent bipolar-like oscillations continue in cases II and III until
$r\gtrsim 250$ km where collective oscillations fade away because of the low neutrino densities.

\begin{figure*}[htb]
  \begin{center}
    $\begin{array}{@{}l@{\hspace{0.01in}}l@{\hspace{0.01in}}l@{}}
      \includegraphics*[scale=\figscalei]{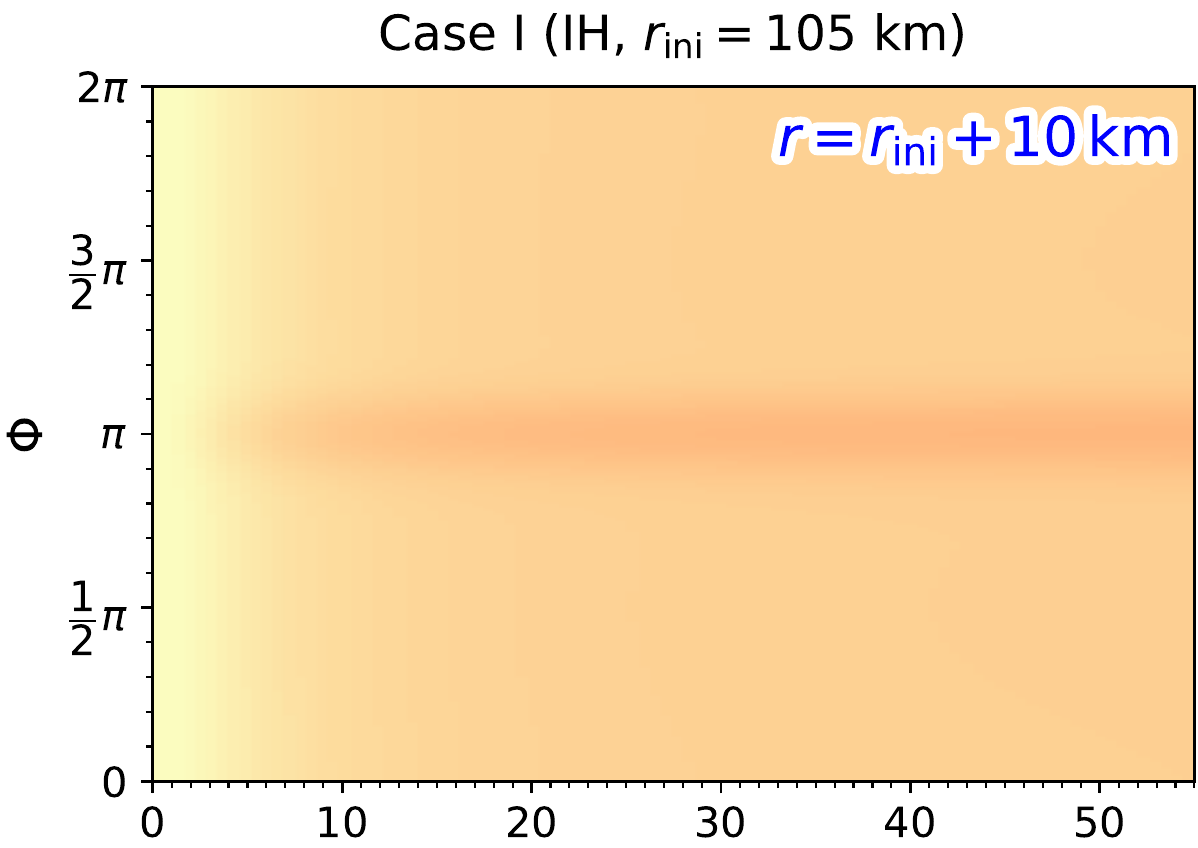} &
      \includegraphics*[scale=\figscalei]{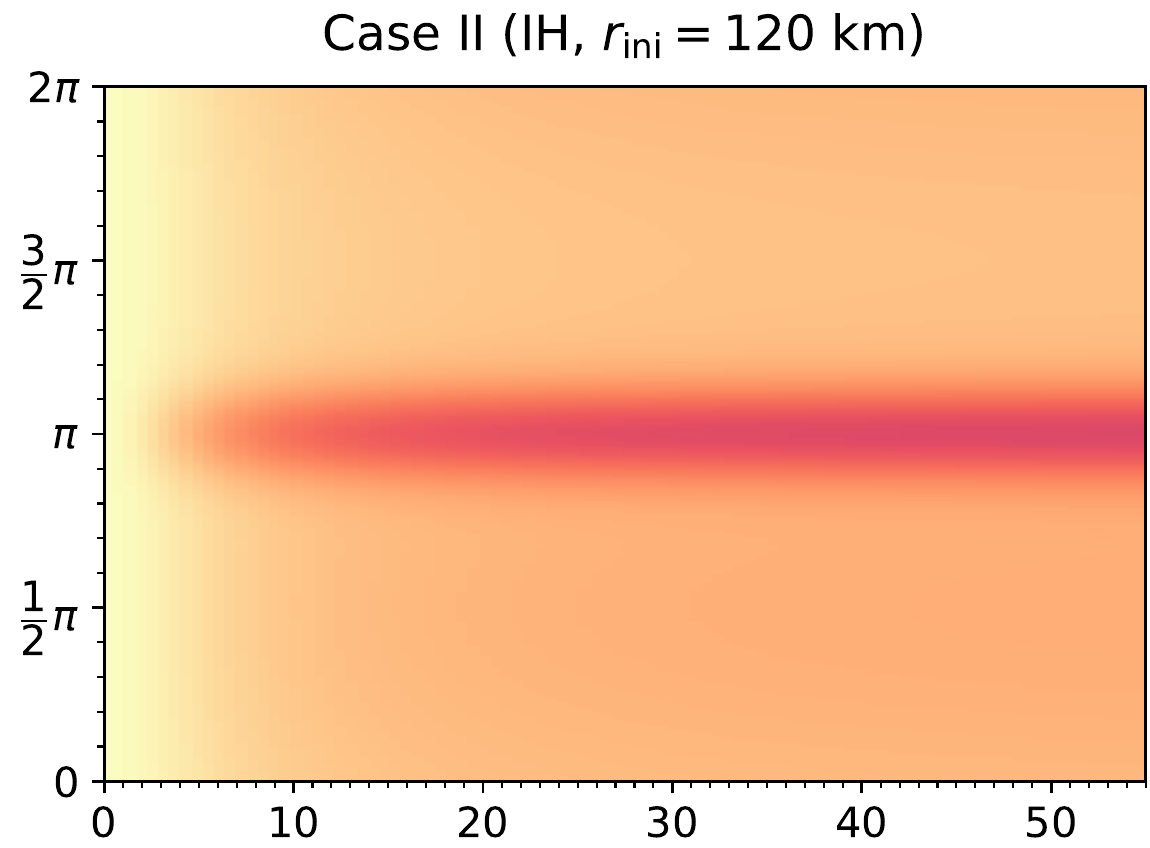} &
      \includegraphics*[scale=\figscalei]{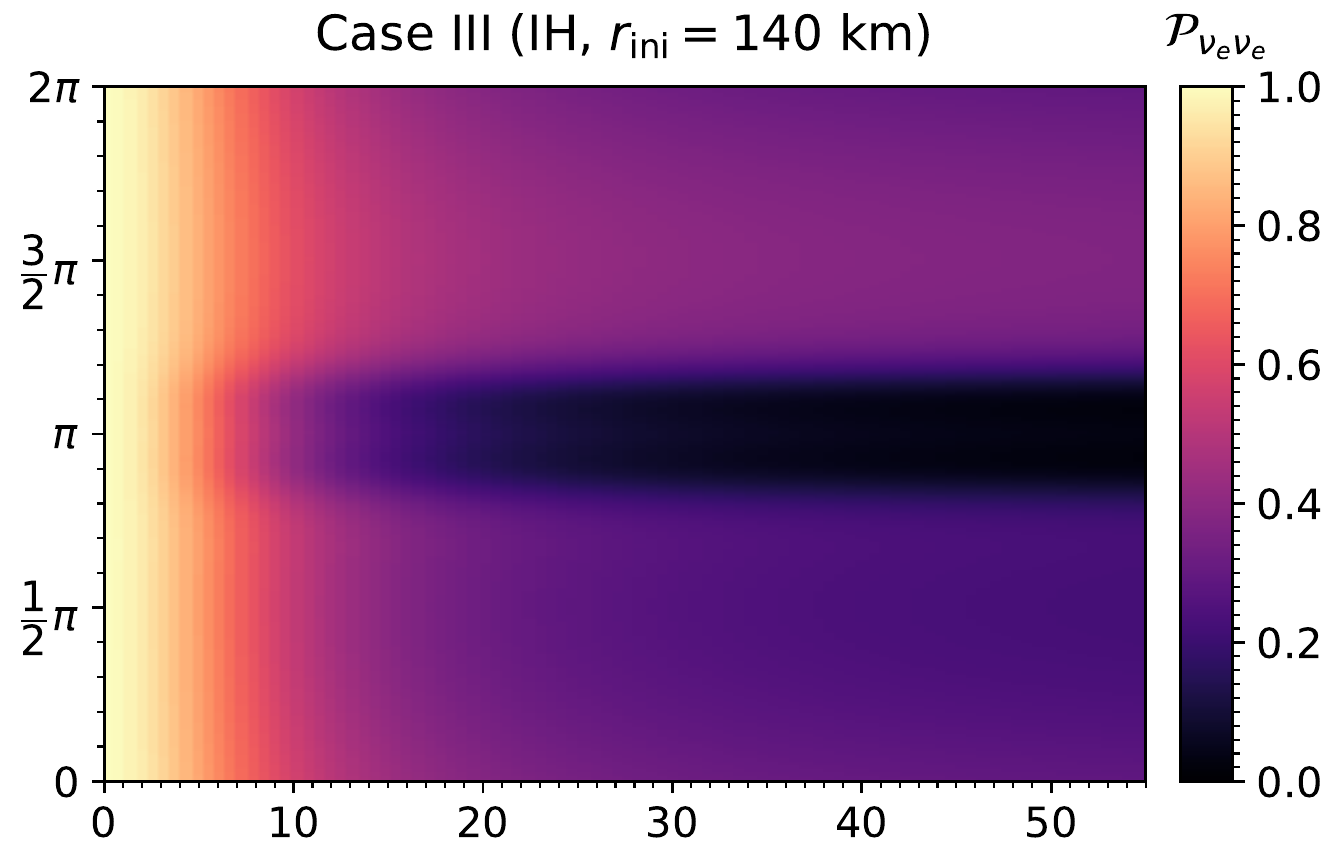} \\
      \includegraphics*[scale=\figscalei]{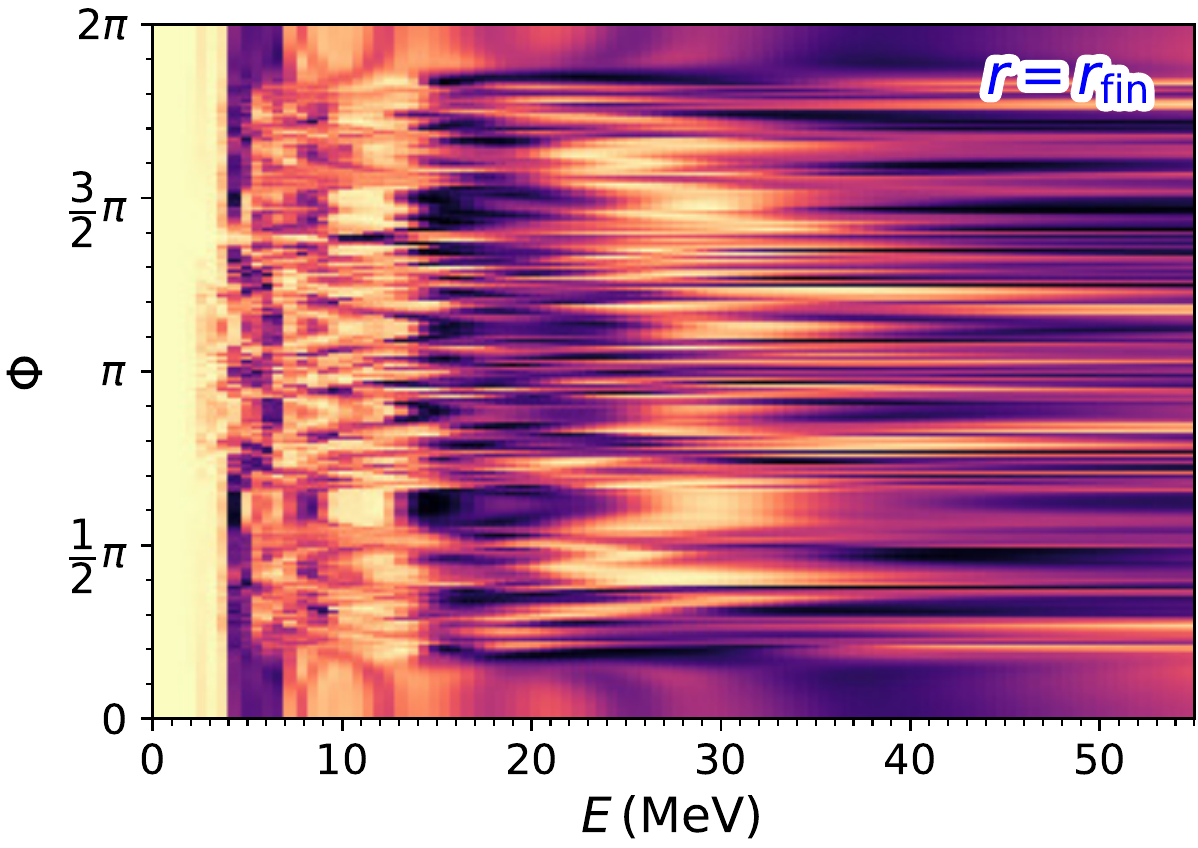} &
      \includegraphics*[scale=\figscalei]{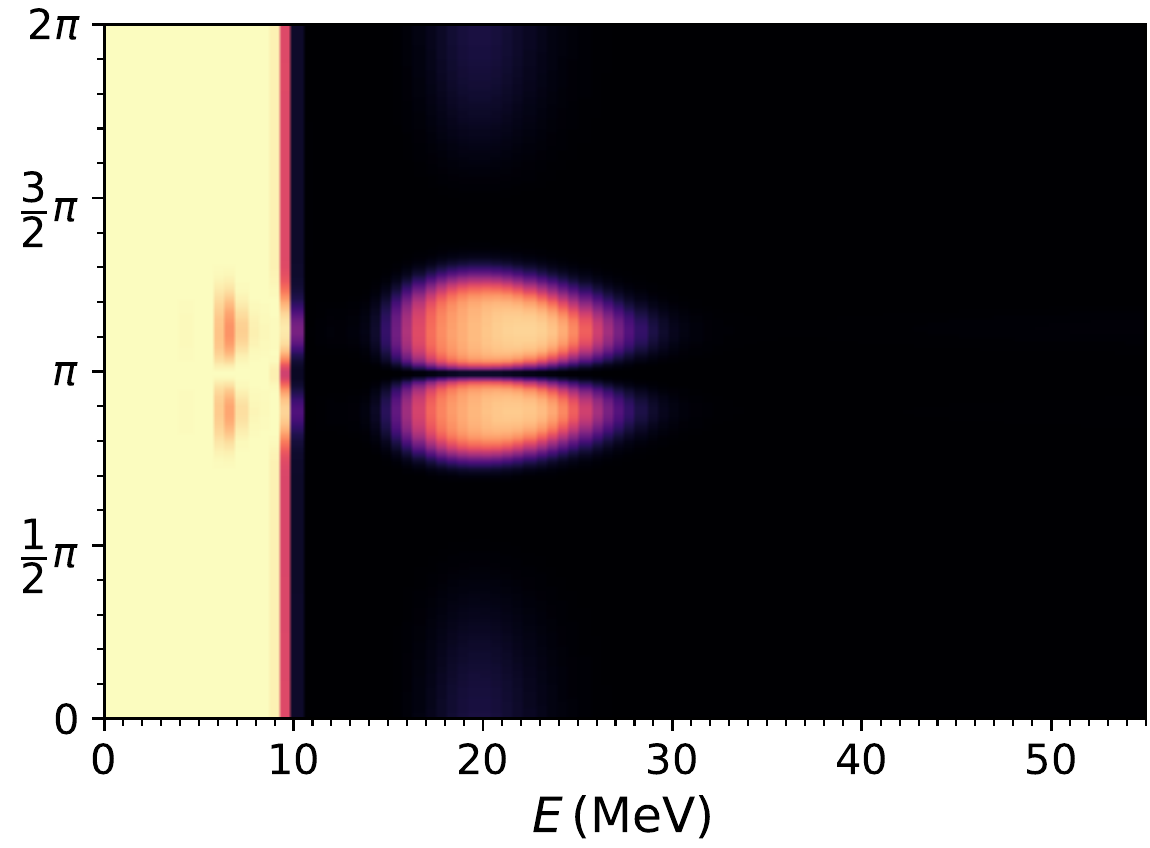} &
      \includegraphics*[scale=\figscalei]{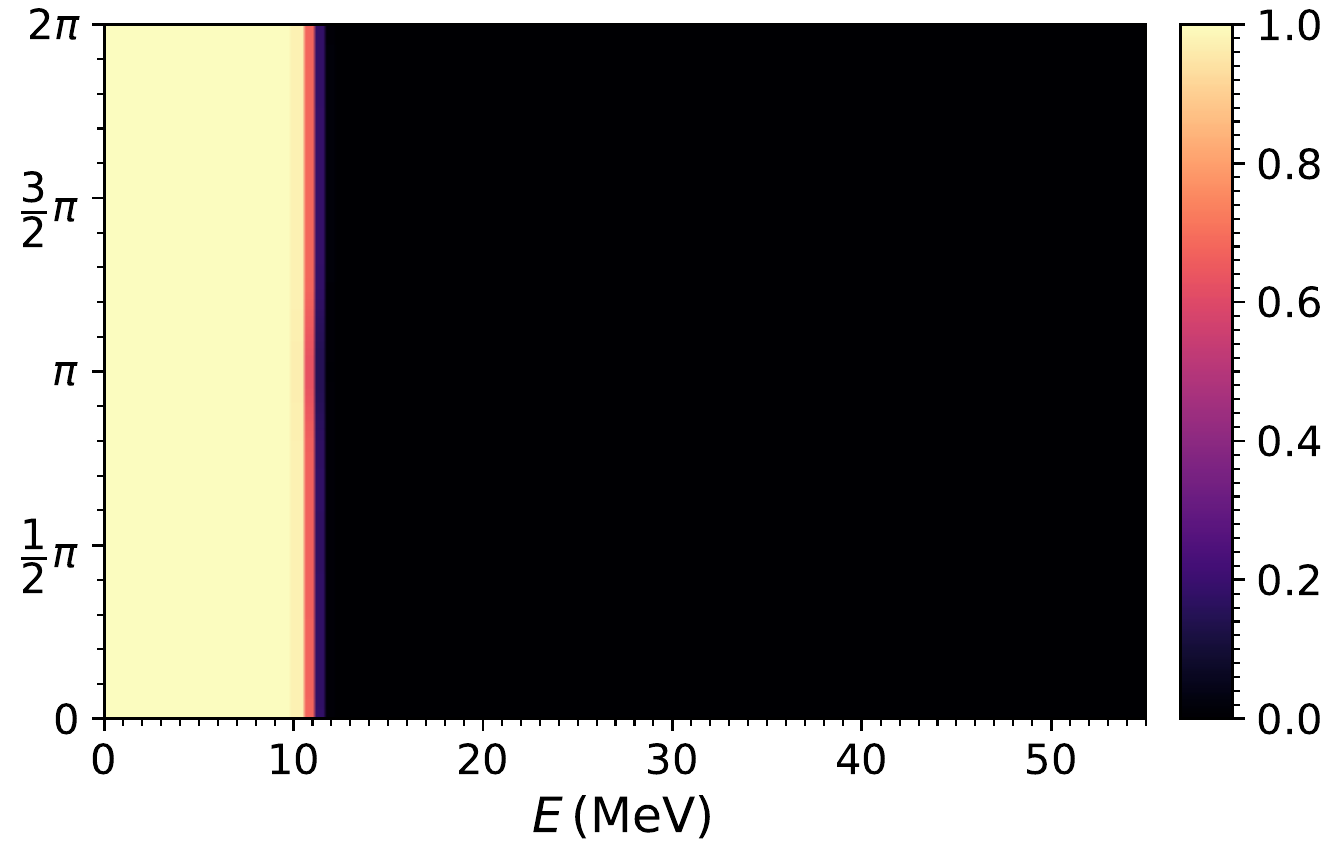}
    \end{array}$
  \end{center}
  \caption{(Color online)
  The $\nu_e$ flavor survival probability $\mathcal{P}_{\nu_e\nu_e}(r, \Phi)$ (averaged over the two neutrino beams) for the same three cases described in Fig.~\ref{fig:nnu-IH} at radii $\rini+10$ km (top panels) and $\rfin=350$ km, respectively.
  }\label{fig:Pnu-IH}
\end{figure*}

In Fig.~\ref{fig:Pnu-IH} we show the electron flavor neutrino survival probabilities $\mathcal{P}_{\nu_e\nu_e}(\Phi, E)$ (averaged over the two neutrino beams) in the above three cases at both 10 km after the calculations begin and the final radius $\rfin=350$ km where we stop the calculations. This figure shows that the bipolar-like oscillations in all three cases are coherent across both the angular space and the energy space initially. It is also clear that the circular symmetry is manifestly broken by collective oscillations in all three cases, especially for neutrinos with energies larger than a few MeV. However, the neutrino gases in the three cases reach different fates at the final radius. In case I, the flavor survival probability $\mathcal{P}_{\nu_e\nu_e}(\Phi, E)$ of the gas at $\rfin$ has rather rapid oscillations in terms of $\Phi$ but a relatively smooth dependence of $E$. In contrast, in case III, the circular symmetry is almost completely restored at $\rfin$ where $\mathcal{P}_{\nu_e\nu_e}(\Phi, E)$
has very little dependence on $\Phi$ and is a step function of $E$. This step-like dependence of  $\mathcal{P}_{\nu\nu}$ is known as the spectral swap/split which is a hallmark of the collective neutrino oscillations in the 1D bulb model \cite{Duan:2006jv,Duan:2006an}.
Case II is in the middle ground between cases I and III. Its final neutrino flavor survival probability $\mathcal{P}_{\nu_e\nu_e}(\Phi, E)$ show both the spectral swaps/splits and explicit breaking of the circular symmetry.

\begin{figure*}[htb]
  \begin{center}
    $\begin{array}{@{}l@{\hspace{0.01in}}l@{\hspace{0.01in}}l@{}}
      \includegraphics*[scale=\figscalei]{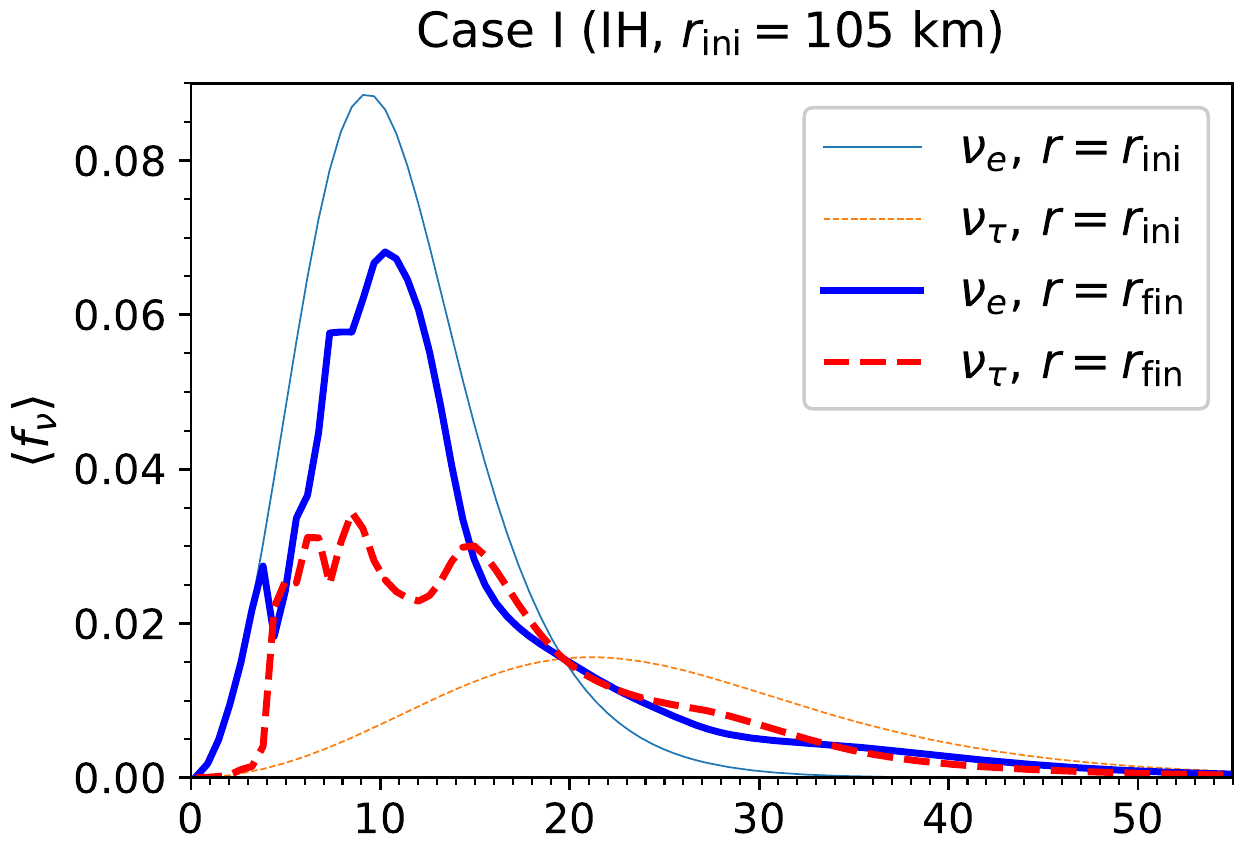} &
      \includegraphics*[scale=\figscalei]{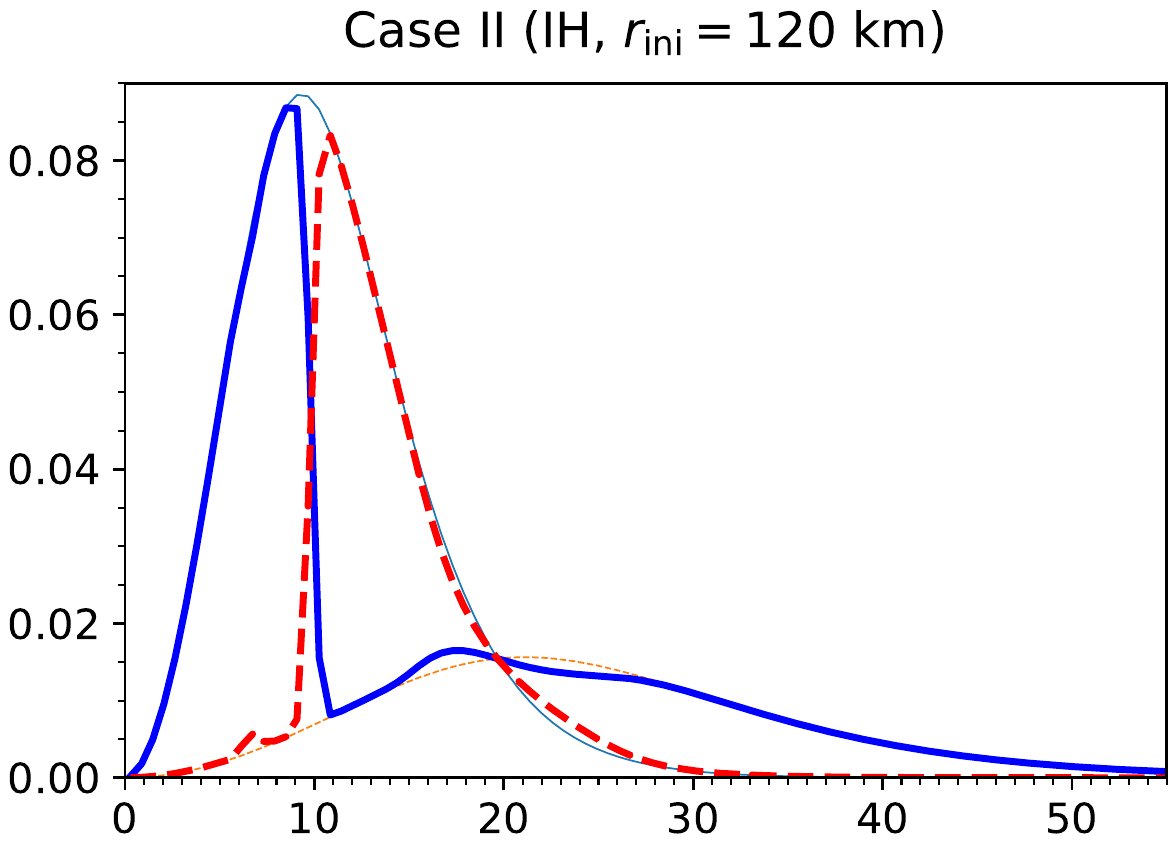} &
      \includegraphics*[scale=\figscalei]{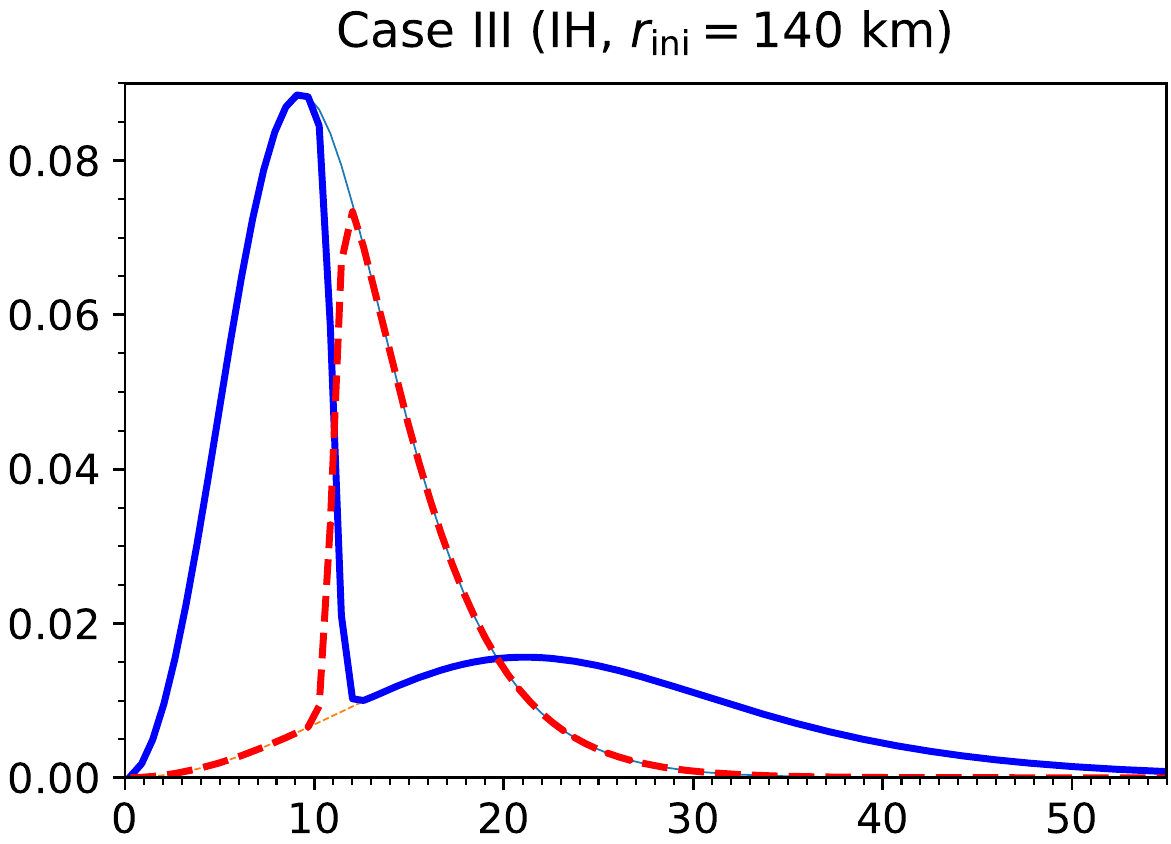} \\
      \includegraphics*[scale=\figscalei]{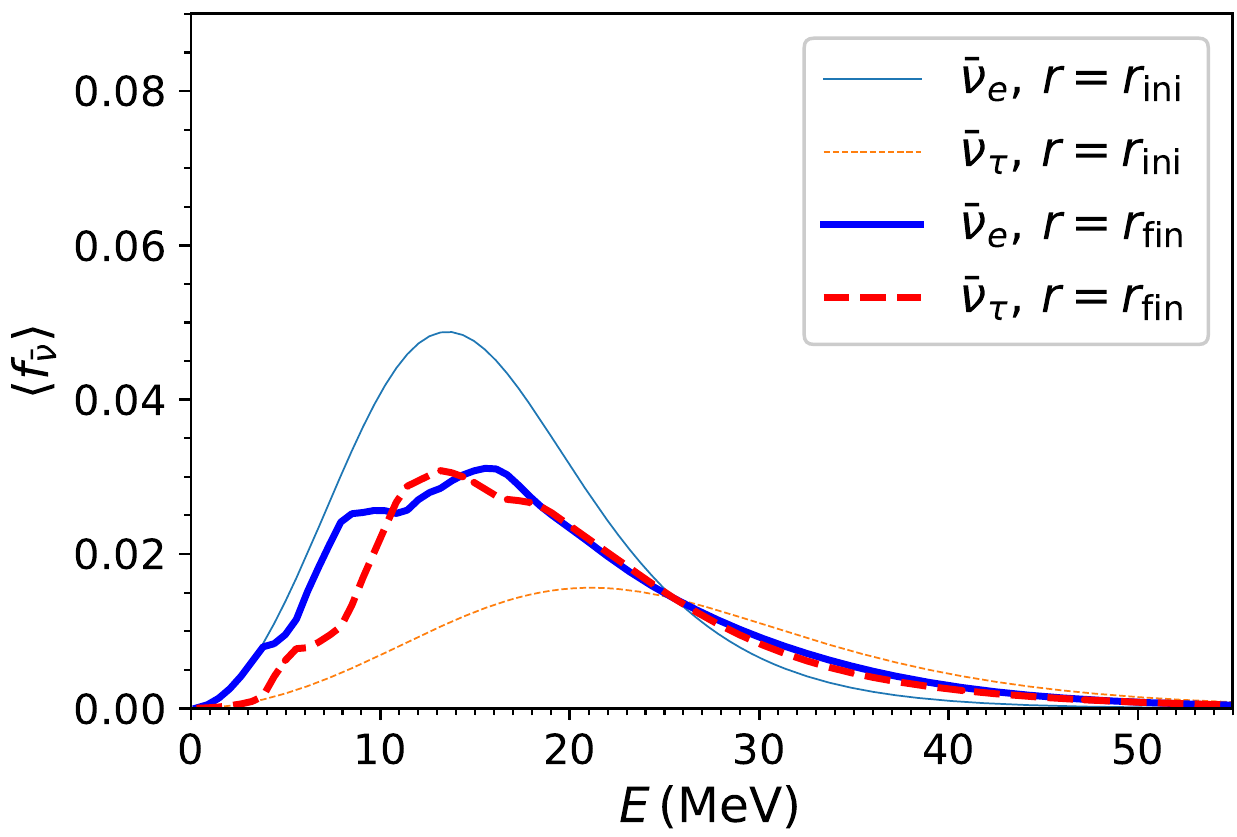} &
      \includegraphics*[scale=\figscalei]{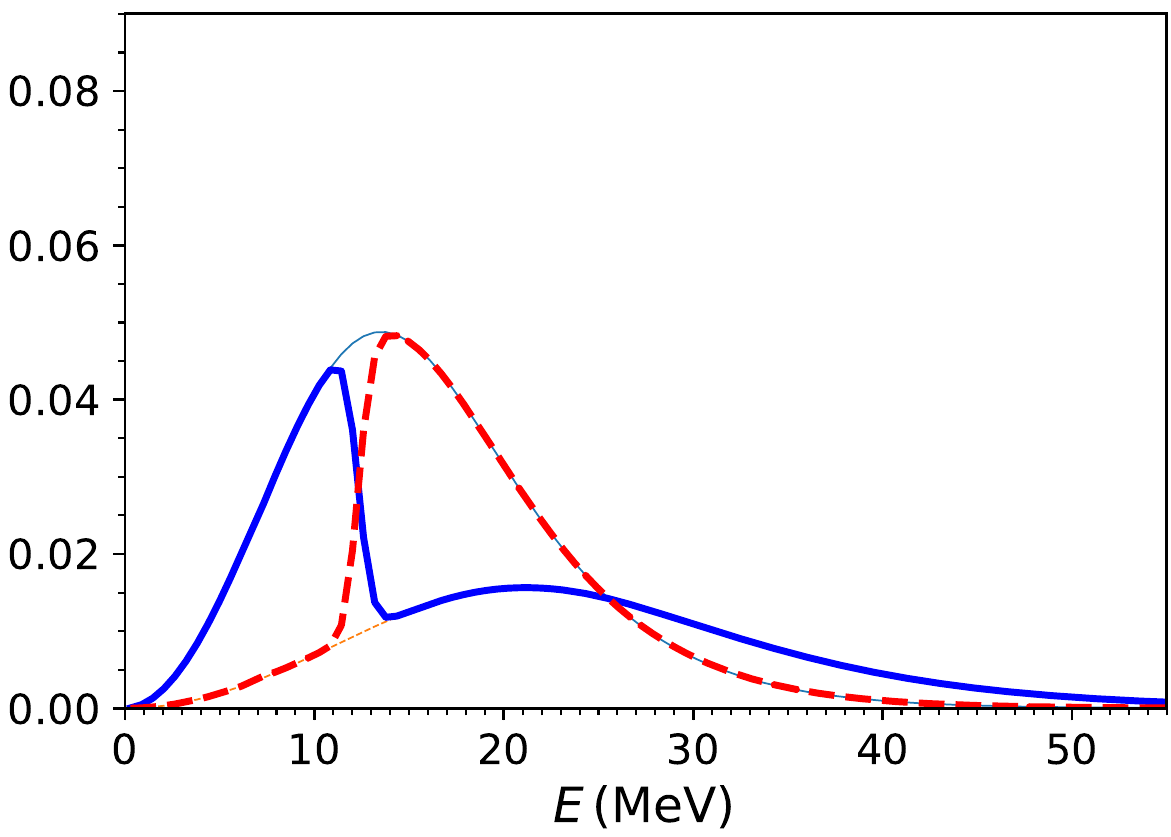} &
      \includegraphics*[scale=\figscalei]{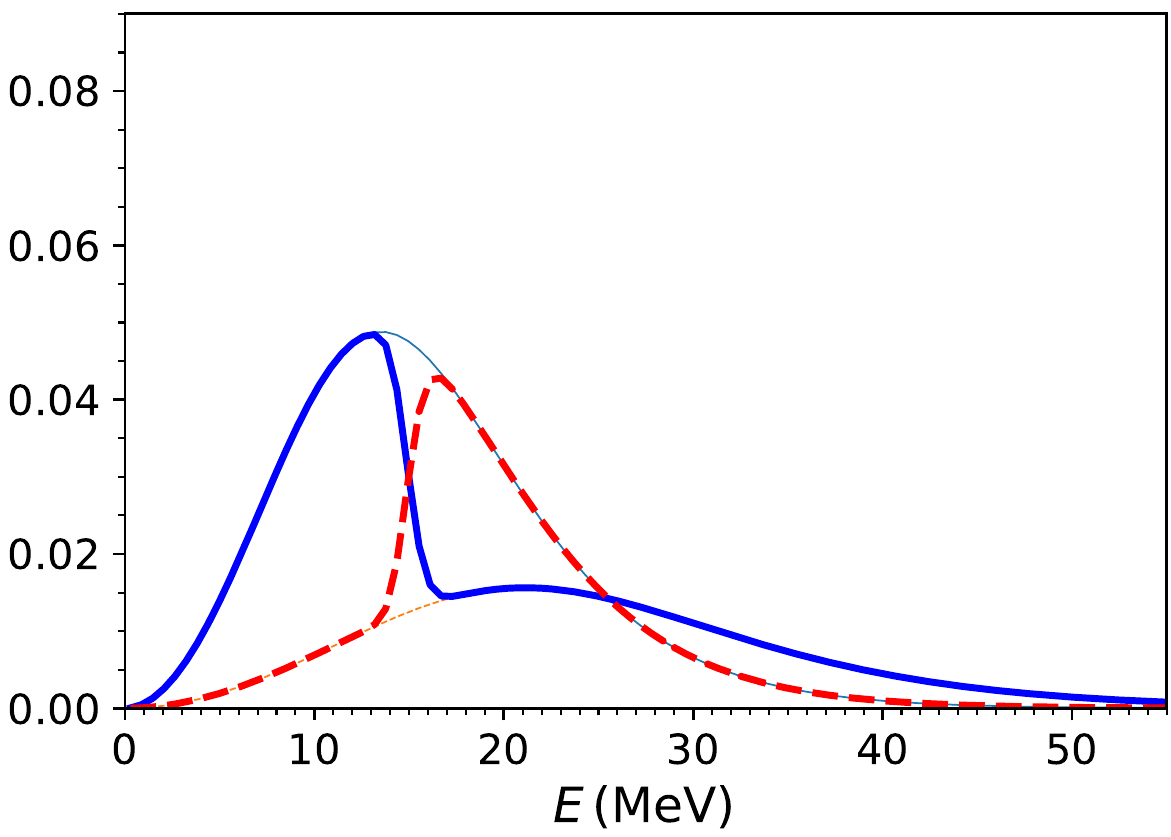}
    \end{array}$
  \end{center}
  \caption{(Color online)
  The average energy spectra $\langle f_\nu(E)\rangle $ (over both the angular coordinate $\Phi$ and the neutrino beams) for various neutrino species and
  at the initial and final radii, $\rini$ and $\rfin$, (as labeled) in the three cases described in Figs.~\ref{fig:nnu-IH} and \ref{fig:Pnu-IH}.
  }\label{fig:fnu-IH}
\end{figure*}

In Fig.~\ref{fig:fnu-IH} we compare the averaged initial and final energy spectra for all neutrino species in cases I through III. In case I, the average spectra of $\bar\nu_e$ and $\bar\nu_\tau$ become similar to each other at $\rfin$, and so are $\nu_e$ and $\nu_\tau$ with $E\gtrsim 15$ MeV. We note that the spectra of $\nu_e$ and $\nu_\tau$ cannot be similar in all energy range, or the ELN $\cal{L}$ defined in Eq.~\eqref{eq:L} would not be constant. In contrast, The $e$ and $\tau$ flavor (anti-) neutrinos partially swap their energy spectra in the other two cases as in the bulb model.

\subsection{Normal mass hierarchy}
\begin{figure*}[htb]
  \begin{center}
    $\begin{array}{@{}l@{\hspace{0.01in}}l@{\hspace{0.01in}}l@{}}
      \includegraphics*[scale=\figscalei]{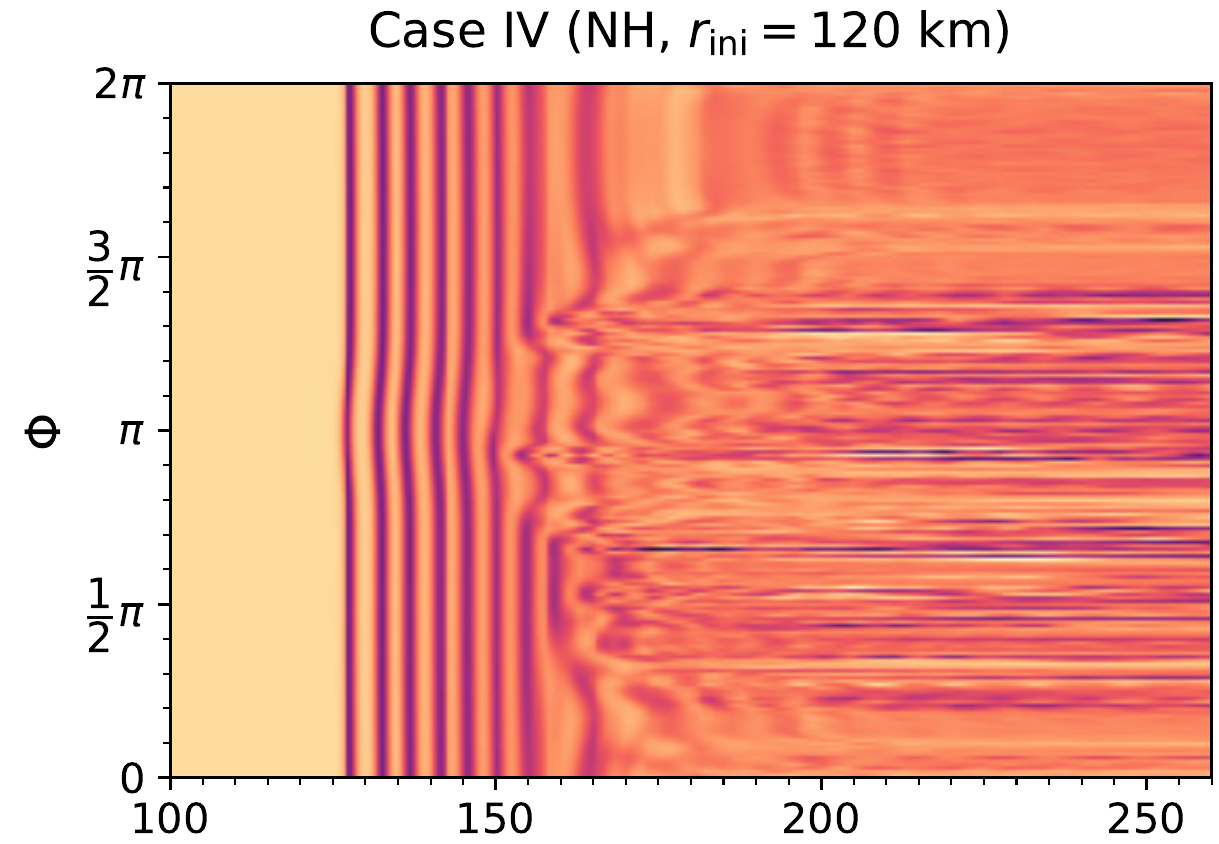} &
      \includegraphics*[scale=\figscalei]{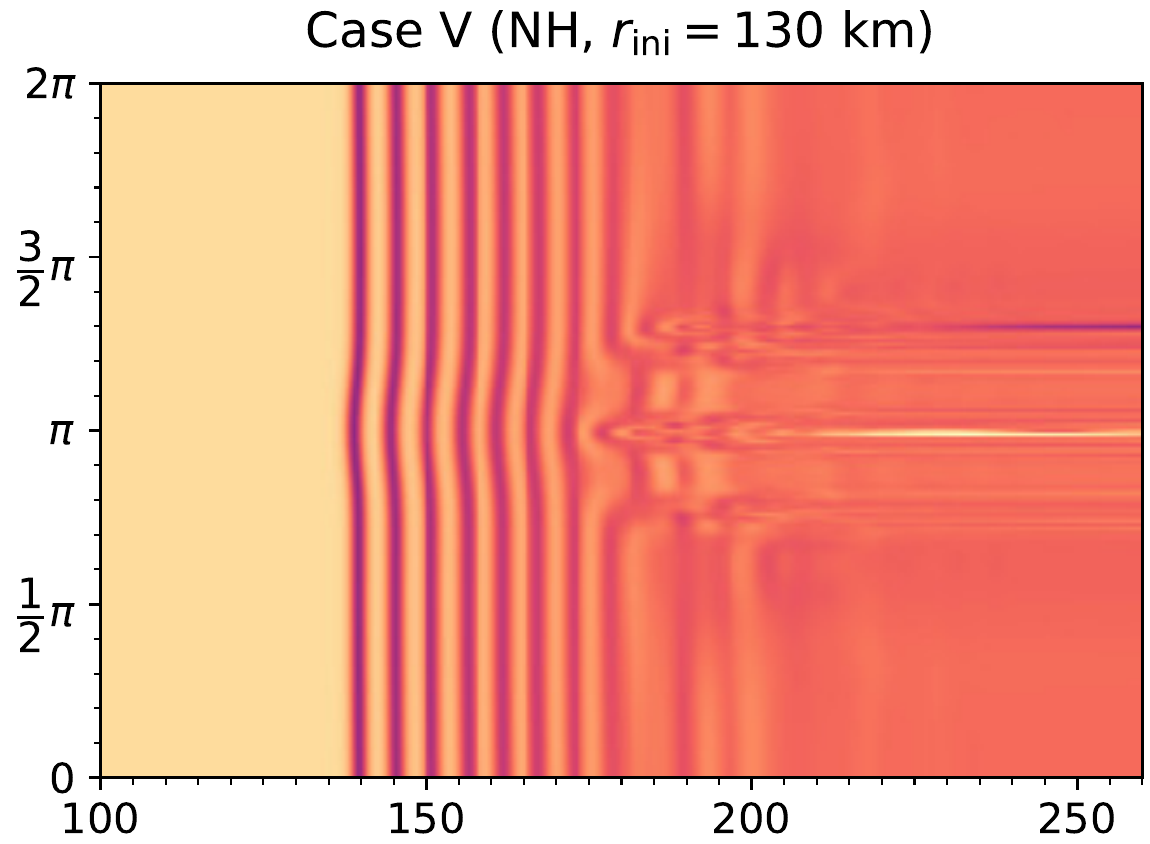} &
      \includegraphics*[scale=\figscalei]{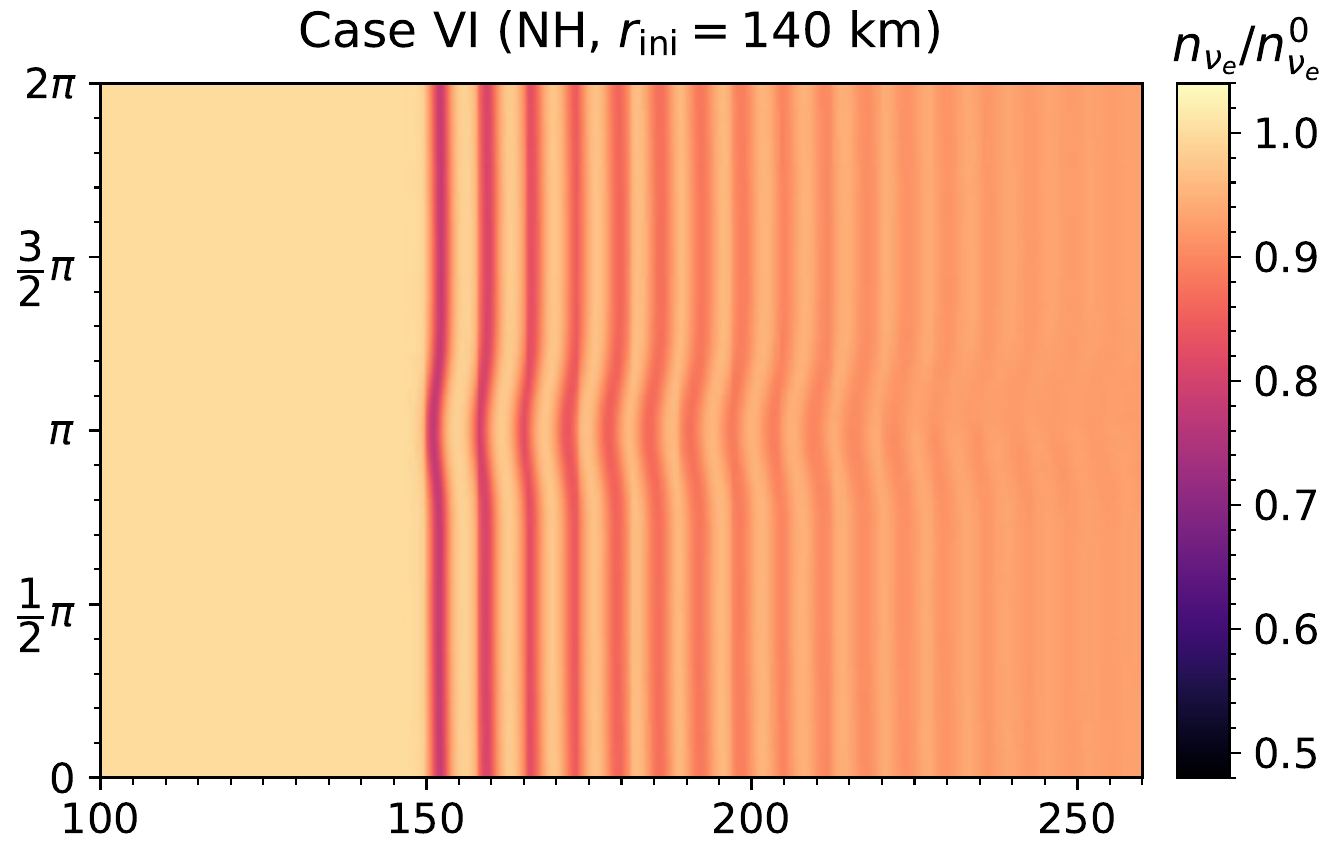} \\
      \includegraphics*[scale=\figscalei]{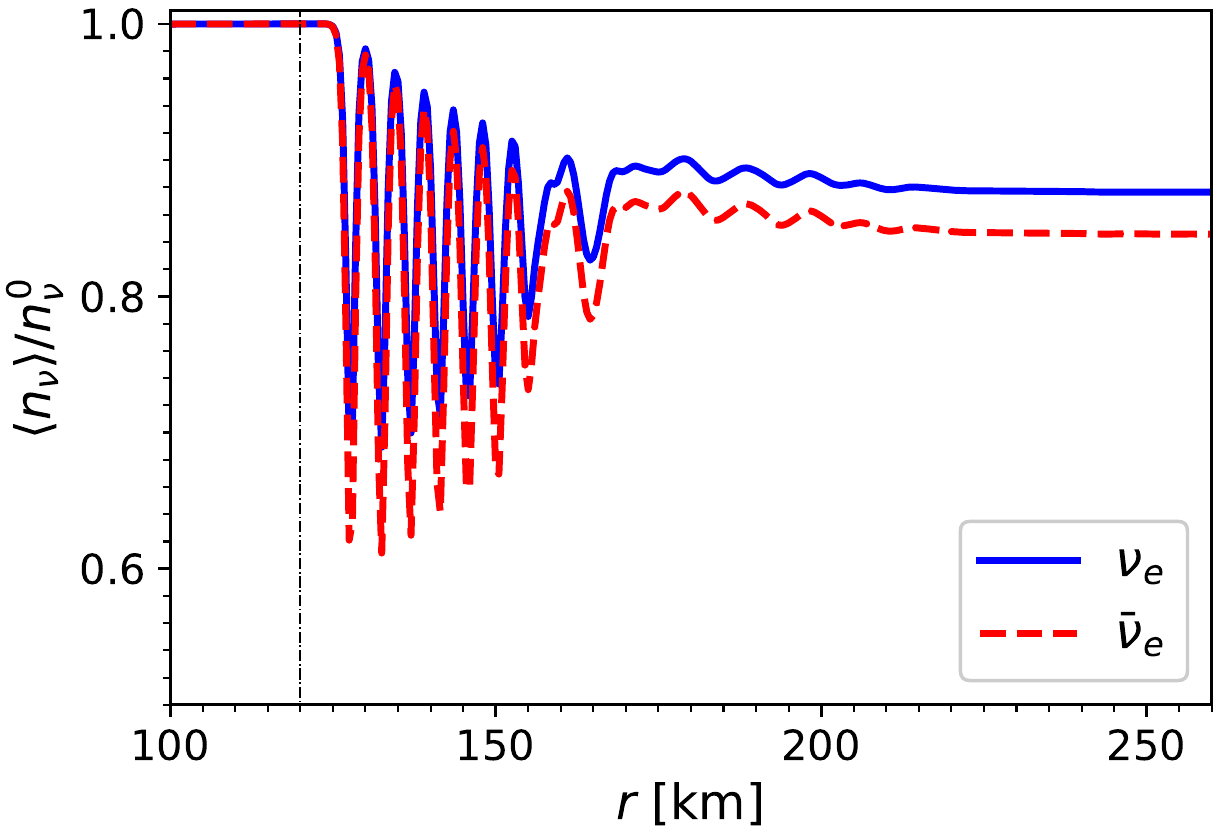} &
      \includegraphics*[scale=\figscalei]{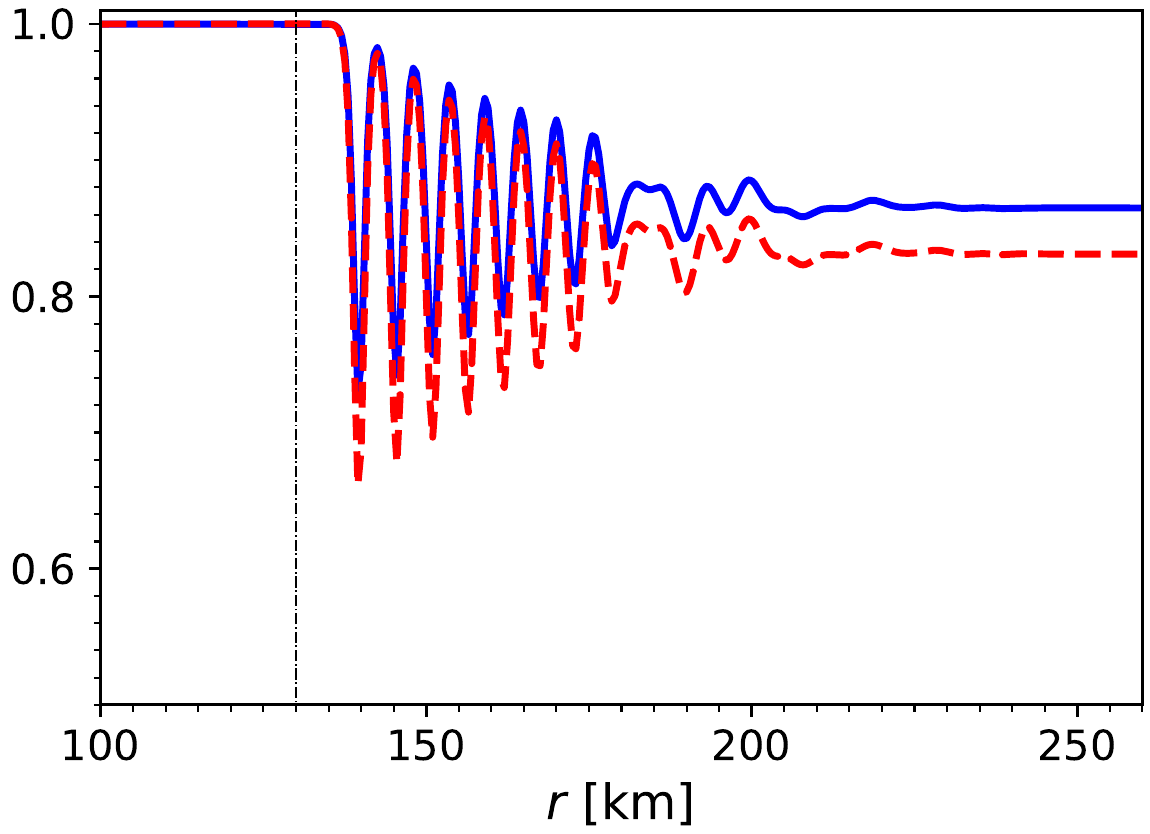} &
      \includegraphics*[scale=\figscalei]{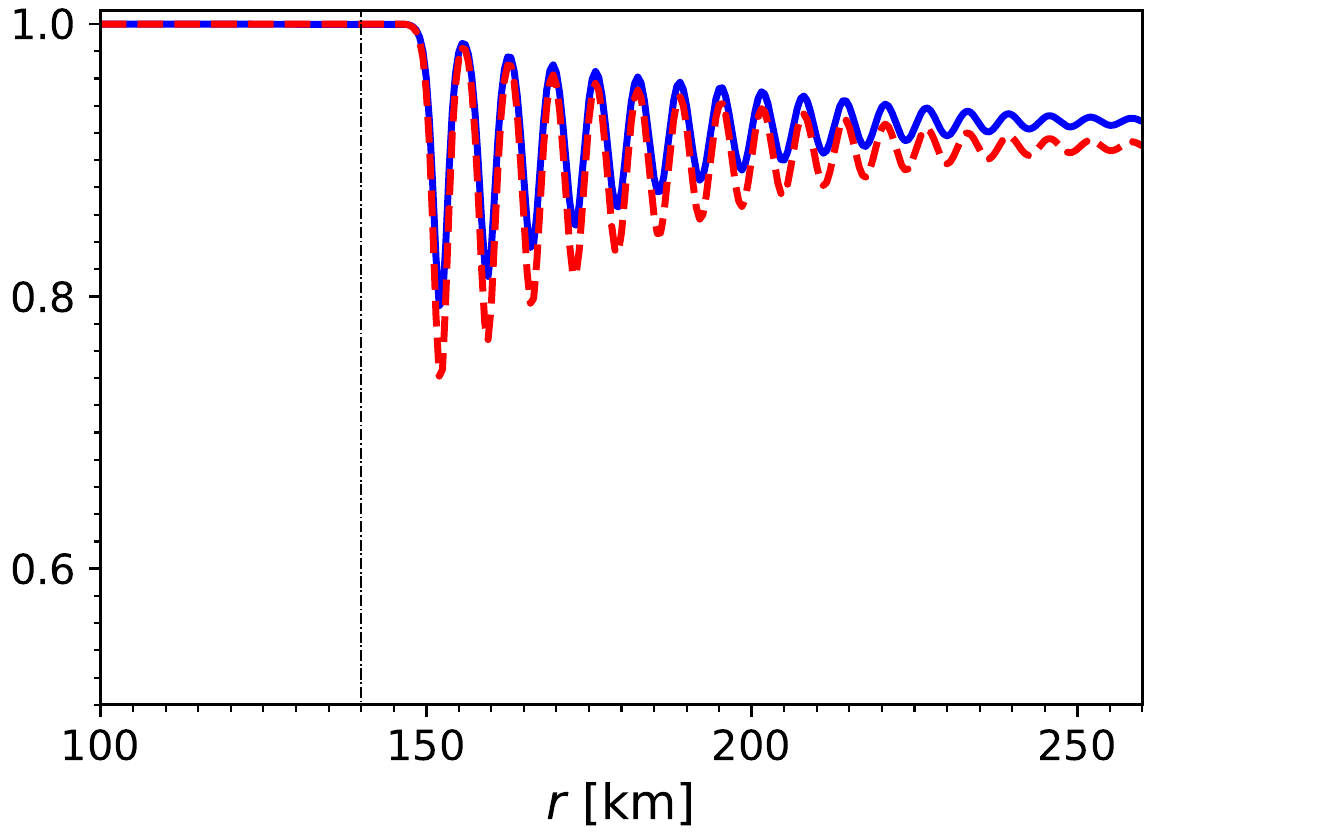}
    \end{array}$
  \end{center}
  \caption{(Color online)
  Similar to Fig.~\ref{fig:nnu-IH} but for a set of 3 calculations, cases IV through VI, with the normal neutrino mass hierarchy (NH) and with $\rini=120$ km, $130$ km, and $140$ km, respectively.
  }\label{fig:nnu-NH}
\end{figure*}

The behaviors of the neutrino gases with the NH are qualitatively the same as those with the IH. In the upper panels of Fig.~\ref{fig:nnu-NH}, we show the ratios $n_{\nu_e}/ n_{\nu_e}^0$ as functions of $r$ and $\Phi$ for cases IV through VI which mimic the scenarios where the matter suppression is lifted at radii $\rini=120$ km, 130 km, and 140 km, respectively. In the lower panels we show the angle averaged ratios $\langle n_{\nu_e} \rangle / n_{\nu_e}^0$ and $\langle n_{\bar\nu_e} \rangle / n_{\bar\nu_e}^0$ as functions of $r$. Compared to the IH cases, the neutrino gases with the NH are more prone to develop fine flavor structures.
Even the neutrino gas in case V with $\rini=130$ km begins to have small-scale flavor structures at $r\approx 180$ km.

\begin{figure*}[htb]
  \begin{center}
    $\begin{array}{@{}l@{\hspace{0.01in}}l@{\hspace{0.01in}}l@{}}
      \includegraphics*[scale=\figscalei]{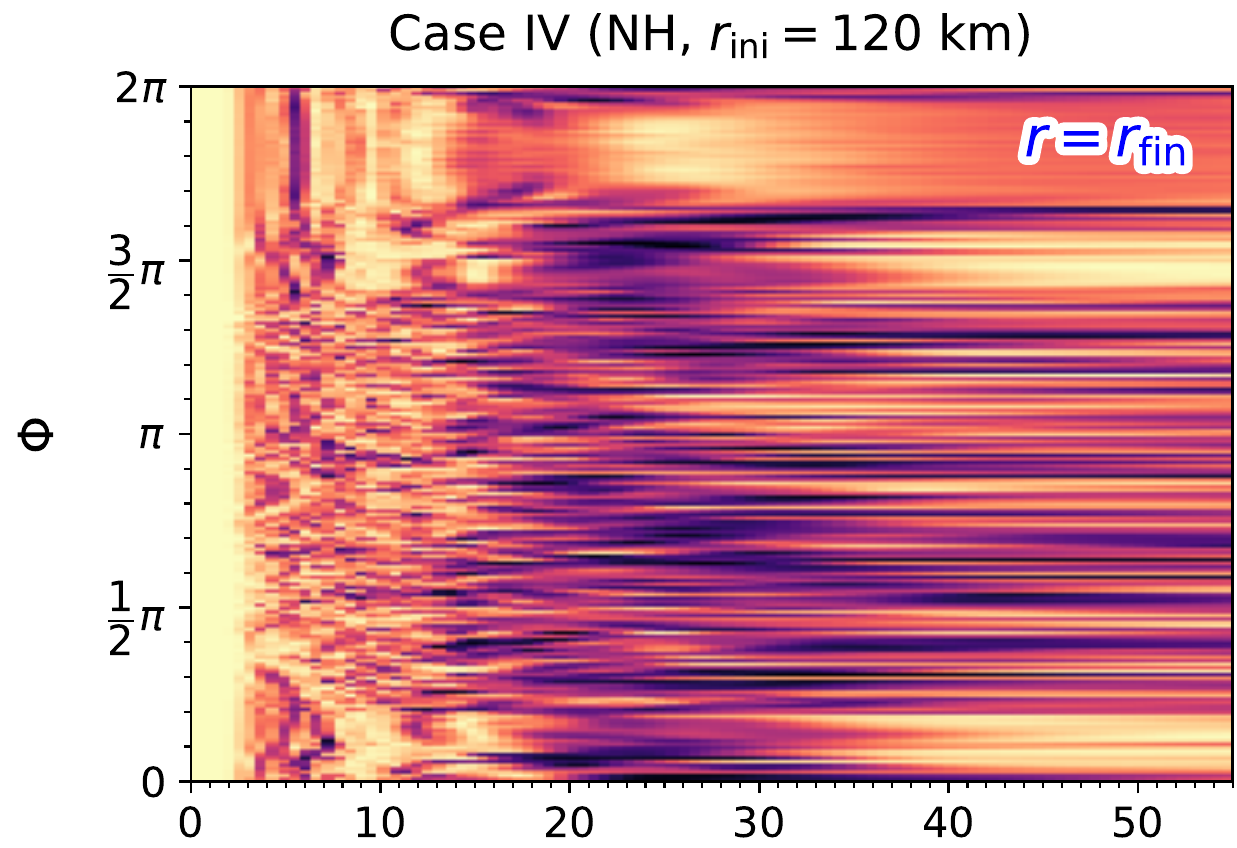} &
      \includegraphics*[scale=\figscalei]{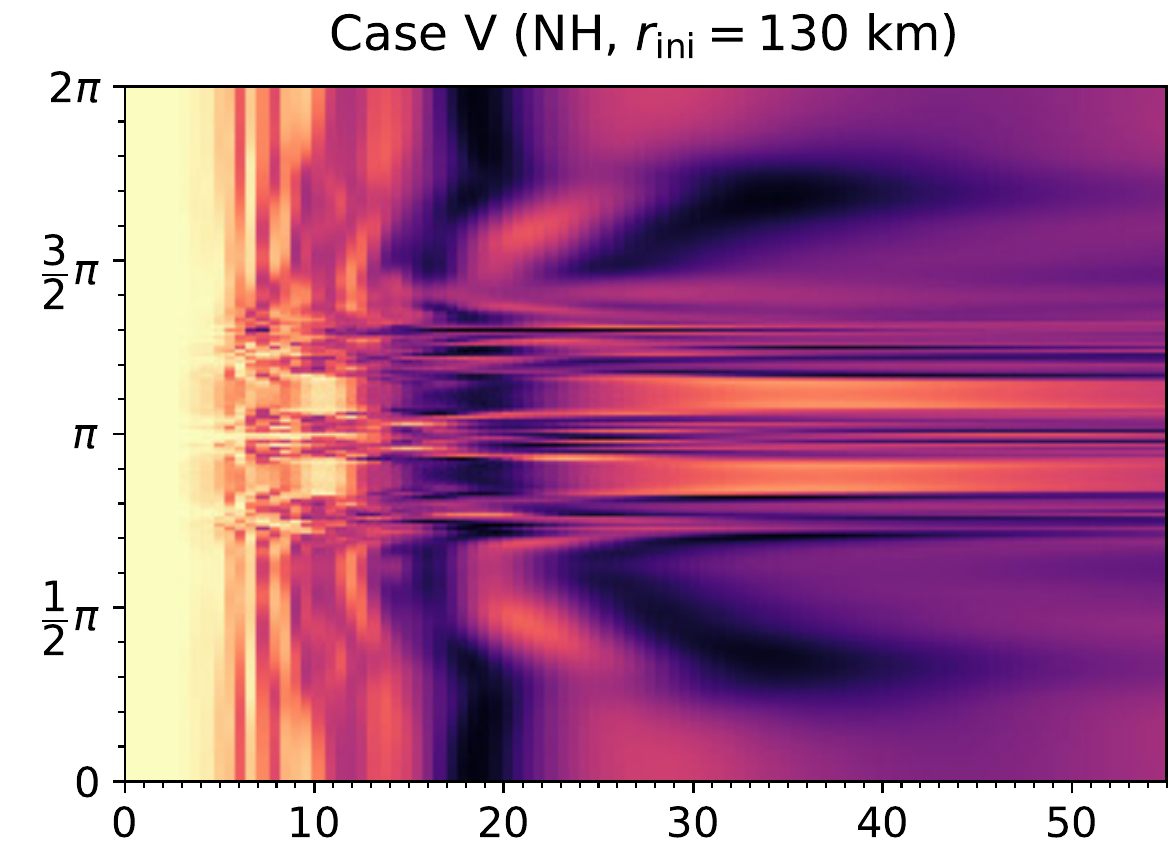} &
      \includegraphics*[scale=\figscalei]{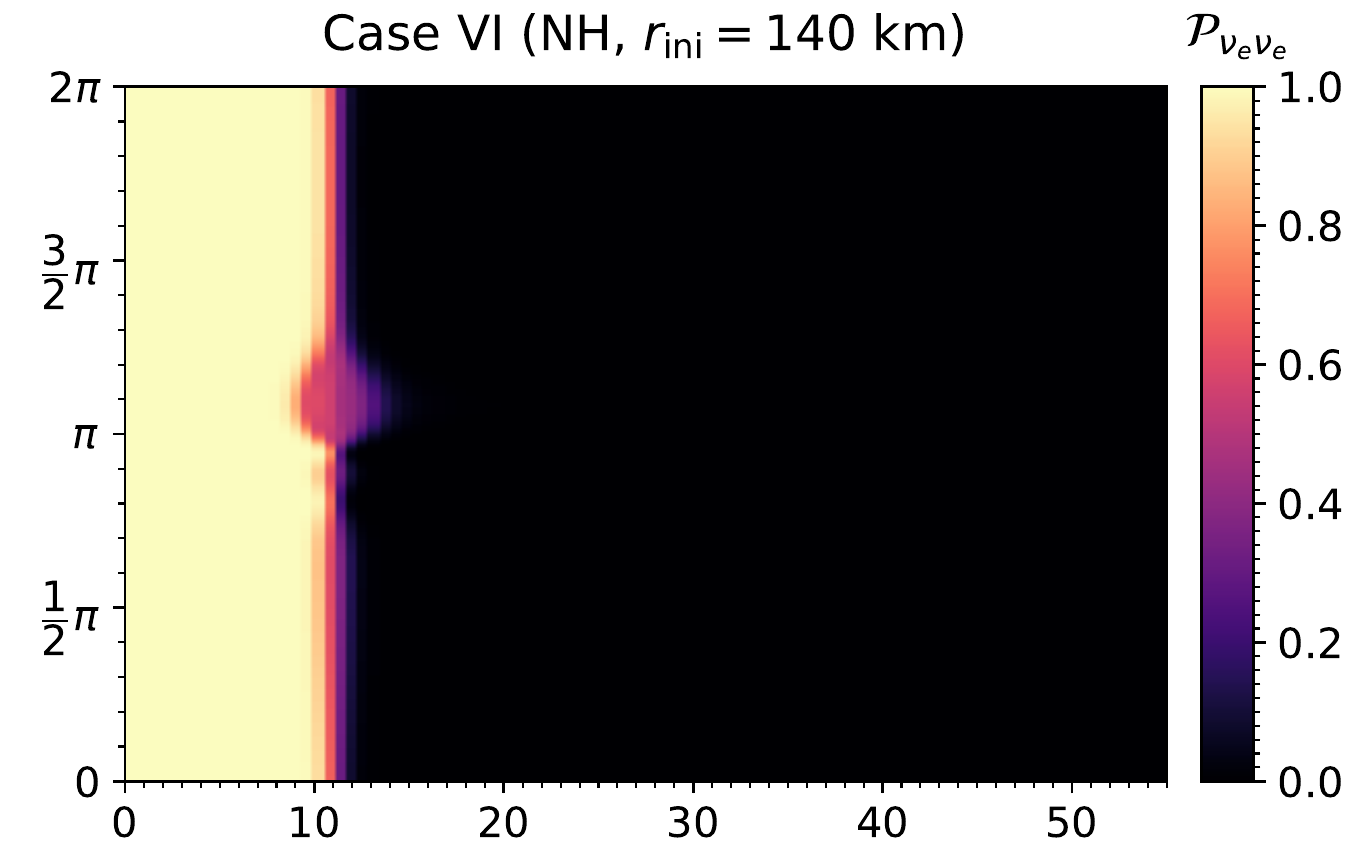} \\
      \includegraphics*[scale=\figscalei]{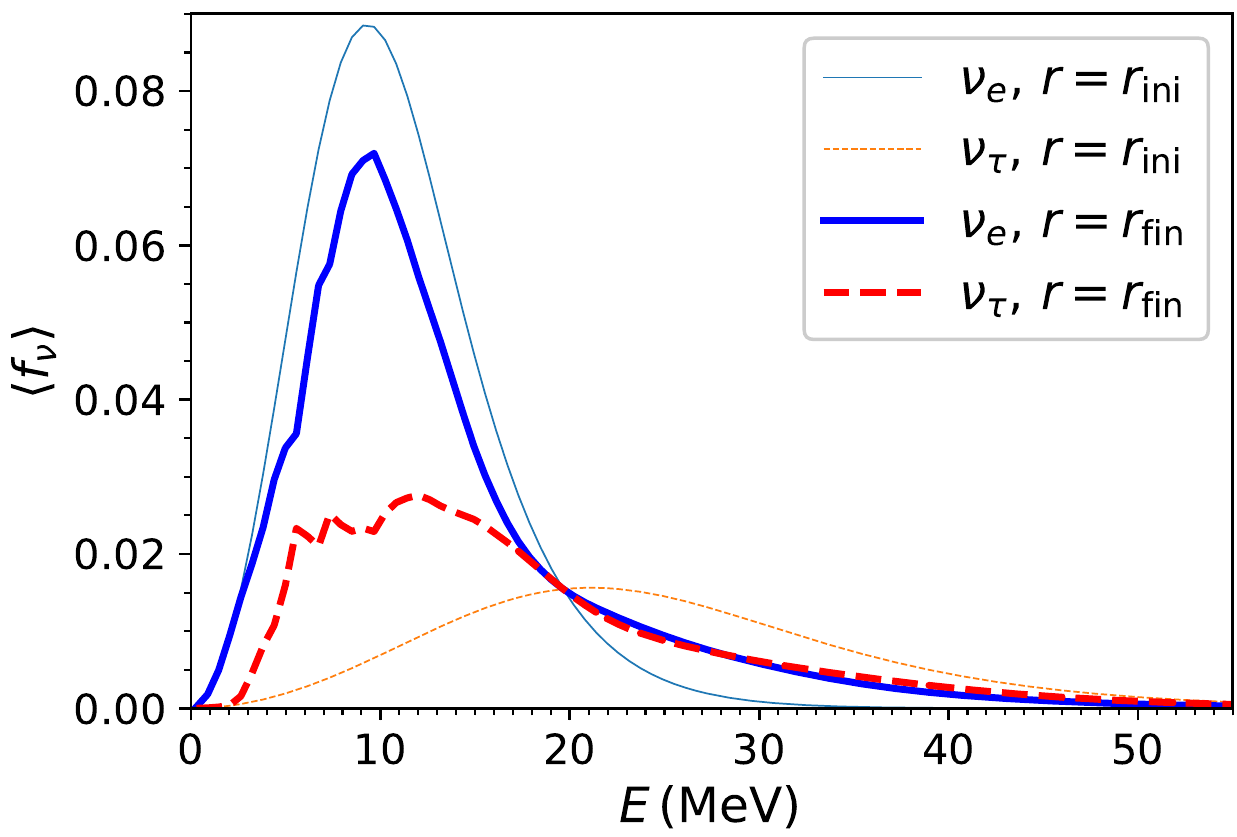} &
      \includegraphics*[scale=\figscalei]{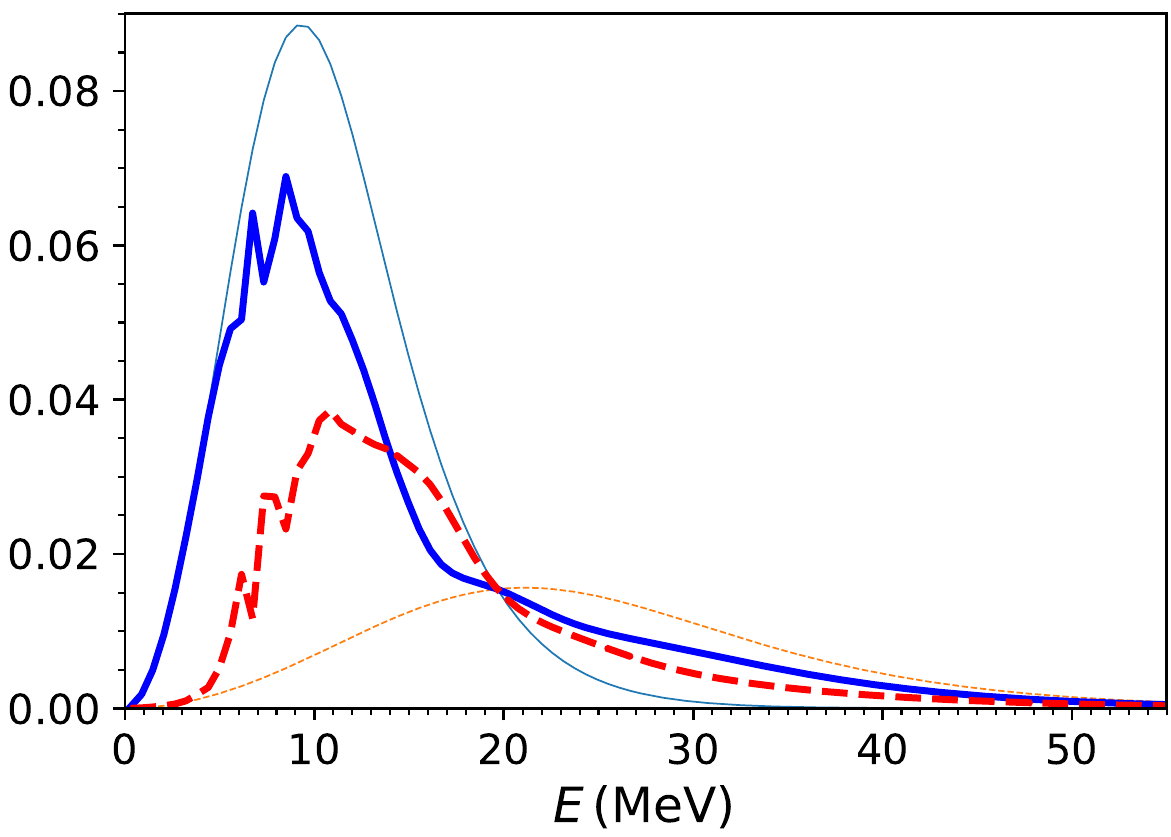} &
      \includegraphics*[scale=\figscalei]{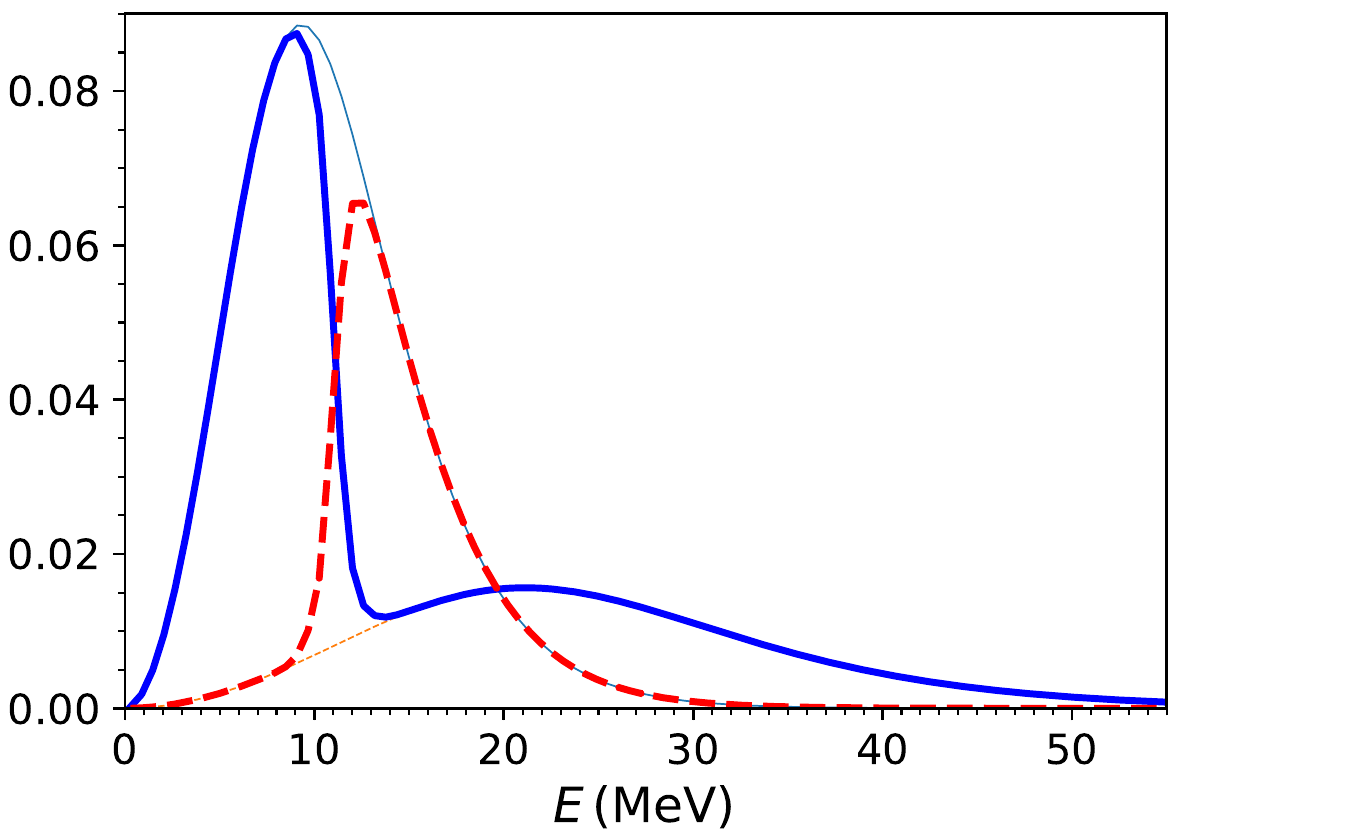}
    \end{array}$
  \end{center}
  \caption{(Color online)
  The $\nu_e$ flavor survival probability $P_{\nu_e\nu_e}(r, \Phi)$ and the average energy spectra $\langle f_\nu(E) \rangle$ for $\nu_e$ and $\nu_\tau$ at $\rfin$ in the three cases described in Fig.~\ref{fig:nnu-NH}.
  }\label{fig:Pf-NH}
\end{figure*}

In Fig.~\ref{fig:Pf-NH} we show both the electron-flavor neutrino survival probabilities and the averaged energy spectra of $\nu_e$ and $\nu_\tau$ at the final radius. Similar to case I, $\mathcal{P}_{\nu_e\nu_e}$ in case IV has rapid oscillations with respect to $\Phi$ but is relatively smooth in $E$. The small-scale flavor structures result in similar $\nu_e$ and $\nu_\tau$ spectra at $E\gtrsim 18$ MeV. Similar to case III, both the survival probability and average energy spectra in case VI have developed spectral swaps/splits. Case V is in the middle ground between cases IV and VI and has more explicit fine flavor structures than case II.

We note that case VI is in contrast to the bulb model which does not produce swap/split in the NH scenario (without a suitable matter profile). This difference is due to the fact that the neutrinos in the bulb model possess an axial symmetry about the radial direction which corresponds to the symmetry between the two neutrino beams in the ring model. The latter symmetry is spontaneously broken when the circular symmetry is violated. It has been shown that the neutrino oscillation modes which break the symmetry in momentum space (e.g., the axial symmetry in the bulb model) in the NH scenarios behave in a way qualitatively similar to the symmetry preserving modes in the corresponding IH scenarios \cite{Raffelt:2013rqa, Mirizzi:2013wda, Raffelt:2013isa, Duan:2013kba}.

\section{Discussion and conclusions} \label{sec:conclusions}

We have developed a numerical code to study the nonlinear behavior of a dense neutrino gas emitted from
a 2D ring.  Inspired by
the suggestion that flavor evolution may be suppressed by a large matter density near the proto-neutron star,
we have elected to investigate the behavior of the gas in the nonlinear regime for cases in which
flavor evolution begins at different radii.  We have chosen a set of Fermi-Dirac energy spectra for the neutrinos of different flavors and
simulated the evolution of the gas from two initial emission angles.

Our results show two distinct behaviors, each of which has been observed in previous studies.
We find that, when flavor conversion begins at higher neutrino number densities (and at lower radii), the spatial correlation of the neutrino flavor field can be largely destroyed. This behavior is similar to that of the neutrino gases in the 2D neutrino line models which have constant neutrino densities. Although the flavor instability window shifts and narrows with the decreasing neutrino densities in the ring model, small-scale flavor structures have been sufficiently developed before the window fully closes.
In this scenario, the antineutrinos of different flavors achieve similar energy spectra in the end. The neutrinos of different flavors with relative high energies also obtain similar spectra, but at low energies the spectra must differ in order to respect the conservation of the ELN.

In our study we also find that there exists an opposite scenario in which
the gas is unstable to self-induced flavor conversion, but the neutrino flavor field remain coherent in space.
In this second scenario, although the spatial symmetry (i.e.\ the circular symmetry in the ring model and the spherical symmetry in the bulb model) is manifestly broken by neutrino oscillations, there is not a sufficient window for the small-scale flavor structures to be fully grown. Remarkably, the spatial symmetry is restored as the collective neutrino oscillations cease.
We also observe spectral swaps with corresponding sharp splits in the survival probabilities.  This spectral swap/split behavior is a characteristic of the neutrino flavor transformation in 1D studies, but until now it has not been observed in 2D models.

We have observed qualitatively similar behaviors for the neutrino gases with the NH and IH, although the NH scenarios are much more prone to the development of small-scale flavor structures than the IH ones. In the limiting scenarios where the spatial correlation of the neutrino flavor field is maintained, the neutrino gases obtain similar spectral swaps in either mass hierarchy. Although we have demonstrated this phenomenon using the energy spectra similar to those with which the spectral swaps were first reported \cite{Duan:2006jv, Duan:2006an}, we have verified that similar results also obtain when other neutrino spectra, e.g., the ones which induce multiple swaps/splits \cite{Dasgupta:2009mg, Duan:2010bf}, are employed.

Our study highlights the inadequacy of the analytic analysis of the flavor stability of the neutrino gas in the linear regime, which does not predict the behavior of the neutrino flavor transformation in the nonlinear regime. Our results suggest that neutrino oscillations, independent of which scenario may occur in a real supernova environment, can have important impacts on the nucleosynthesis and the neutrino signals (e.g., see Refs.~\cite{Duan:2010af, Wu:2014kaa, Mirizzi:2015eza}).

The neutrino ring model we adopted has only two neutrino beams from each emitting point. It is known that the dispersion in the flavor oscillations of the neutrinos propagating in different trajectories can introduce kinematic decoherence and even suppress oscillations \cite{Raffelt:2007yz, EstebanPretel:2007ec, Duan:2010bf}. We have artificially begun our calculations at different radii to mimic the suppression of the large matter density. These shortcomings will be addressed in future works.

\section*{Acknowledgements}

We thank S.~Abbar, E.~Grohs, A.~Lovato, S.~T.~Omanakuttan, and S.~Shalgar for useful discussions.  This work is supported
by the U.S.\ DOE Office of Science Graduate Student Research (SCGSR) program (J.D.M.), DOE Office of Science NP Grant No.~DE-SC0017803 at UNM (H.D.\ and J.D.M.), and DOE Office of Science NP under contract AC52-06NA25396 (J.C.).
This research used the resources of the National Energy Research Scientific Computing Center (NERSC), a U.S.\ Department of Energy Office of Science User Facility operated under Contract No.\ DE-AC02-05CH11231. 
The SCGSR program is administered by the Oak Ridge Institute for Science and Education (ORISE) for the DOE. ORISE is managed by ORAU under contract number DE‐SC0014664. All opinions expressed in this paper are the author's and do not necessarily reflect the policies and views of DOE, ORAU, or ORISE.

\bibliography{ringModel}

\begin{thebibliography}{40}%
\makeatletter
\providecommand \@ifxundefined [1]{%
 \@ifx{#1\undefined}
}%
\providecommand \@ifnum [1]{%
 \ifnum #1\expandafter \@firstoftwo
 \else \expandafter \@secondoftwo
 \fi
}%
\providecommand \@ifx [1]{%
 \ifx #1\expandafter \@firstoftwo
 \else \expandafter \@secondoftwo
 \fi
}%
\providecommand \natexlab [1]{#1}%
\providecommand \enquote  [1]{``#1''}%
\providecommand \bibnamefont  [1]{#1}%
\providecommand \bibfnamefont [1]{#1}%
\providecommand \citenamefont [1]{#1}%
\providecommand \href@noop [0]{\@secondoftwo}%
\providecommand \href [0]{\begingroup \@sanitize@url \@href}%
\providecommand \@href[1]{\@@startlink{#1}\@@href}%
\providecommand \@@href[1]{\endgroup#1\@@endlink}%
\providecommand \@sanitize@url [0]{\catcode `\\12\catcode `\$12\catcode
  `\&12\catcode `\#12\catcode `\^12\catcode `\_12\catcode `\%12\relax}%
\providecommand \@@startlink[1]{}%
\providecommand \@@endlink[0]{}%
\providecommand \url  [0]{\begingroup\@sanitize@url \@url }%
\providecommand \@url [1]{\endgroup\@href {#1}{\urlprefix }}%
\providecommand \urlprefix  [0]{URL }%
\providecommand \Eprint [0]{\href }%
\providecommand \doibase [0]{http://dx.doi.org/}%
\providecommand \selectlanguage [0]{\@gobble}%
\providecommand \bibinfo  [0]{\@secondoftwo}%
\providecommand \bibfield  [0]{\@secondoftwo}%
\providecommand \translation [1]{[#1]}%
\providecommand \BibitemOpen [0]{}%
\providecommand \bibitemStop [0]{}%
\providecommand \bibitemNoStop [0]{.\EOS\space}%
\providecommand \EOS [0]{\spacefactor3000\relax}%
\providecommand \BibitemShut  [1]{\csname bibitem#1\endcsname}%
\let\auto@bib@innerbib\@empty
\bibitem [{\citenamefont {Tanabashi}\ \emph {et~al.}(2018)\citenamefont
  {Tanabashi} \emph {et~al.}}]{Tanabashi:2018oca}%
  \BibitemOpen
  \bibfield  {author} {\bibinfo {author} {\bibfnamefont {M.}~\bibnamefont
  {Tanabashi}} \emph {et~al.} (\bibinfo {collaboration} {Particle Data
  Group}),\ }\bibfield  {title} {\enquote {\bibinfo {title} {{Review of
  Particle Physics}},}\ }\href {\doibase 10.1103/PhysRevD.98.030001} {\bibfield
   {journal} {\bibinfo  {journal} {Phys. Rev.}\ }\textbf {\bibinfo {volume}
  {D98}},\ \bibinfo {pages} {030001} (\bibinfo {year} {2018})}\BibitemShut
  {NoStop}%
\bibitem [{\citenamefont {Mikheyev}\ and\ \citenamefont
  {Smirnov}(1985)}]{Mikheev:1986gs}%
  \BibitemOpen
  \bibfield  {author} {\bibinfo {author} {\bibfnamefont {S.~P.}\ \bibnamefont
  {Mikheyev}}\ and\ \bibinfo {author} {\bibfnamefont {A.}~\bibnamefont
  {Smirnov}},\ }\bibfield  {title} {\enquote {\bibinfo {title} {{Resonance
  Amplification of Oscillations in Matter and Spectroscopy of Solar
  Neutrinos}},}\ }\href@noop {} {\bibfield  {journal} {\bibinfo  {journal}
  {Sov. J. Nucl. Phys.}\ }\textbf {\bibinfo {volume} {42}},\ \bibinfo {pages}
  {913--917} (\bibinfo {year} {1985})},\ \bibinfo {note}
  {[,305(1986)]}\BibitemShut {NoStop}%
\bibitem [{\citenamefont {Wolfenstein}(1978)}]{Wolfenstein:1977ue}%
  \BibitemOpen
  \bibfield  {author} {\bibinfo {author} {\bibfnamefont {L.}~\bibnamefont
  {Wolfenstein}},\ }\bibfield  {title} {\enquote {\bibinfo {title} {{Neutrino
  Oscillations in Matter}},}\ }\href {\doibase 10.1103/PhysRevD.17.2369}
  {\bibfield  {journal} {\bibinfo  {journal} {Phys. Rev.}\ }\textbf {\bibinfo
  {volume} {D17}},\ \bibinfo {pages} {2369--2374} (\bibinfo {year} {1978})},\
  \bibinfo {note} {[,294(1977)]}\BibitemShut {NoStop}%
\bibitem [{\citenamefont {Wolfenstein}(1979)}]{Wolfenstein:1979ni}%
  \BibitemOpen
  \bibfield  {author} {\bibinfo {author} {\bibfnamefont {L.}~\bibnamefont
  {Wolfenstein}},\ }\bibfield  {title} {\enquote {\bibinfo {title} {{Neutrino
  Oscillations and Stellar Collapse}},}\ }\href {\doibase
  10.1103/PhysRevD.20.2634} {\bibfield  {journal} {\bibinfo  {journal} {Phys.
  Rev.}\ }\textbf {\bibinfo {volume} {D20}},\ \bibinfo {pages} {2634--2635}
  (\bibinfo {year} {1979})}\BibitemShut {NoStop}%
\bibitem [{\citenamefont {{Fuller}}\ \emph {et~al.}(1987)\citenamefont
  {{Fuller}}, \citenamefont {{Mayle}}, \citenamefont {{Wilson}},\ and\
  \citenamefont {{Schramm}}}]{1987ApJ...322..795F}%
  \BibitemOpen
  \bibfield  {author} {\bibinfo {author} {\bibfnamefont {G.~M.}\ \bibnamefont
  {{Fuller}}}, \bibinfo {author} {\bibfnamefont {R.~W.}\ \bibnamefont
  {{Mayle}}}, \bibinfo {author} {\bibfnamefont {J.~R.}\ \bibnamefont
  {{Wilson}}}, \ and\ \bibinfo {author} {\bibfnamefont {D.~N.}\ \bibnamefont
  {{Schramm}}},\ }\bibfield  {title} {\enquote {\bibinfo {title} {{Resonant
  neutrino oscillations and stellar collapse}},}\ }\href {\doibase
  10.1086/165772} {\bibfield  {journal} {\bibinfo  {journal} {\apj}\ }\textbf
  {\bibinfo {volume} {322}},\ \bibinfo {pages} {795--803} (\bibinfo {year}
  {1987})}\BibitemShut {NoStop}%
\bibitem [{\citenamefont {Notzold}\ and\ \citenamefont
  {Raffelt}(1988)}]{Notzold:1987ik}%
  \BibitemOpen
  \bibfield  {author} {\bibinfo {author} {\bibfnamefont {Dirk}\ \bibnamefont
  {Notzold}}\ and\ \bibinfo {author} {\bibfnamefont {Georg}\ \bibnamefont
  {Raffelt}},\ }\bibfield  {title} {\enquote {\bibinfo {title} {{Neutrino
  Dispersion at Finite Temperature and Density}},}\ }\href {\doibase
  10.1016/0550-3213(88)90113-7} {\bibfield  {journal} {\bibinfo  {journal}
  {Nucl. Phys.}\ }\textbf {\bibinfo {volume} {B307}},\ \bibinfo {pages}
  {924--936} (\bibinfo {year} {1988})}\BibitemShut {NoStop}%
\bibitem [{\citenamefont {Pantaleone}(1992)}]{Pantaleone:1992xh}%
  \BibitemOpen
  \bibfield  {author} {\bibinfo {author} {\bibfnamefont {James~T.}\
  \bibnamefont {Pantaleone}},\ }\bibfield  {title} {\enquote {\bibinfo {title}
  {{Dirac neutrinos in dense matter}},}\ }\href {\doibase
  10.1103/PhysRevD.46.510} {\bibfield  {journal} {\bibinfo  {journal} {Phys.
  Rev.}\ }\textbf {\bibinfo {volume} {D46}},\ \bibinfo {pages} {510--523}
  (\bibinfo {year} {1992})}\BibitemShut {NoStop}%
\bibitem [{\citenamefont {Kostelecky}\ and\ \citenamefont
  {Samuel}(1995)}]{Kostelecky:1994dt}%
  \BibitemOpen
  \bibfield  {author} {\bibinfo {author} {\bibfnamefont {V.~Alan}\ \bibnamefont
  {Kostelecky}}\ and\ \bibinfo {author} {\bibfnamefont {Stuart}\ \bibnamefont
  {Samuel}},\ }\bibfield  {title} {\enquote {\bibinfo {title} {{Selfmaintained
  coherent oscillations in dense neutrino gases}},}\ }\href {\doibase
  10.1103/PhysRevD.52.621} {\bibfield  {journal} {\bibinfo  {journal} {Phys.
  Rev.}\ }\textbf {\bibinfo {volume} {D52}},\ \bibinfo {pages} {621--627}
  (\bibinfo {year} {1995})},\ \Eprint {http://arxiv.org/abs/hep-ph/9506262}
  {arXiv:hep-ph/9506262 [hep-ph]} \BibitemShut {NoStop}%
\bibitem [{\citenamefont {Duan}\ \emph
  {et~al.}(2006{\natexlab{a}})\citenamefont {Duan}, \citenamefont {Fuller},\
  and\ \citenamefont {Qian}}]{Duan:2005cp}%
  \BibitemOpen
  \bibfield  {author} {\bibinfo {author} {\bibfnamefont {Huaiyu}\ \bibnamefont
  {Duan}}, \bibinfo {author} {\bibfnamefont {George~M.}\ \bibnamefont
  {Fuller}}, \ and\ \bibinfo {author} {\bibfnamefont {Yong-Zhong}\ \bibnamefont
  {Qian}},\ }\bibfield  {title} {\enquote {\bibinfo {title} {{Collective
  neutrino flavor transformation in supernovae}},}\ }\href {\doibase
  10.1103/PhysRevD.74.123004} {\bibfield  {journal} {\bibinfo  {journal} {Phys.
  Rev.}\ }\textbf {\bibinfo {volume} {D74}},\ \bibinfo {pages} {123004}
  (\bibinfo {year} {2006}{\natexlab{a}})},\ \Eprint
  {http://arxiv.org/abs/astro-ph/0511275} {arXiv:astro-ph/0511275 [astro-ph]}
  \BibitemShut {NoStop}%
\bibitem [{\citenamefont {Hannestad}\ \emph {et~al.}(2006)\citenamefont
  {Hannestad}, \citenamefont {Raffelt}, \citenamefont {Sigl},\ and\
  \citenamefont {Wong}}]{Hannestad:2006nj}%
  \BibitemOpen
  \bibfield  {author} {\bibinfo {author} {\bibfnamefont {Steen}\ \bibnamefont
  {Hannestad}}, \bibinfo {author} {\bibfnamefont {Georg~G.}\ \bibnamefont
  {Raffelt}}, \bibinfo {author} {\bibfnamefont {Gunter}\ \bibnamefont {Sigl}},
  \ and\ \bibinfo {author} {\bibfnamefont {Yvonne Y.~Y.}\ \bibnamefont
  {Wong}},\ }\bibfield  {title} {\enquote {\bibinfo {title} {{Self-induced
  conversion in dense neutrino gases: Pendulum in flavour space}},}\ }\href
  {\doibase 10.1103/PhysRevD.74.105010, 10.1103/PhysRevD.76.029901} {\bibfield
  {journal} {\bibinfo  {journal} {Phys. Rev.}\ }\textbf {\bibinfo {volume}
  {D74}},\ \bibinfo {pages} {105010} (\bibinfo {year} {2006})},\ \bibinfo
  {note} {[Erratum: Phys. Rev.D76,029901(2007)]},\ \Eprint
  {http://arxiv.org/abs/astro-ph/0608695} {arXiv:astro-ph/0608695 [astro-ph]}
  \BibitemShut {NoStop}%
\bibitem [{\citenamefont {Duan}\ \emph
  {et~al.}(2006{\natexlab{b}})\citenamefont {Duan}, \citenamefont {Fuller},
  \citenamefont {Carlson},\ and\ \citenamefont {Qian}}]{Duan:2006an}%
  \BibitemOpen
  \bibfield  {author} {\bibinfo {author} {\bibfnamefont {Huaiyu}\ \bibnamefont
  {Duan}}, \bibinfo {author} {\bibfnamefont {George~M.}\ \bibnamefont
  {Fuller}}, \bibinfo {author} {\bibfnamefont {J}~\bibnamefont {Carlson}}, \
  and\ \bibinfo {author} {\bibfnamefont {Yong-Zhong}\ \bibnamefont {Qian}},\
  }\bibfield  {title} {\enquote {\bibinfo {title} {{Simulation of Coherent
  Non-Linear Neutrino Flavor Transformation in the Supernova Environment. 1.
  Correlated Neutrino Trajectories}},}\ }\href {\doibase
  10.1103/PhysRevD.74.105014} {\bibfield  {journal} {\bibinfo  {journal} {Phys.
  Rev.}\ }\textbf {\bibinfo {volume} {D74}},\ \bibinfo {pages} {105014}
  (\bibinfo {year} {2006}{\natexlab{b}})},\ \Eprint
  {http://arxiv.org/abs/astro-ph/0606616} {arXiv:astro-ph/0606616 [astro-ph]}
  \BibitemShut {NoStop}%
\bibitem [{\citenamefont {Duan}\ \emph
  {et~al.}(2006{\natexlab{c}})\citenamefont {Duan}, \citenamefont {Fuller},
  \citenamefont {Carlson},\ and\ \citenamefont {Qian}}]{Duan:2006jv}%
  \BibitemOpen
  \bibfield  {author} {\bibinfo {author} {\bibfnamefont {Huaiyu}\ \bibnamefont
  {Duan}}, \bibinfo {author} {\bibfnamefont {George~M.}\ \bibnamefont
  {Fuller}}, \bibinfo {author} {\bibfnamefont {J.}~\bibnamefont {Carlson}}, \
  and\ \bibinfo {author} {\bibfnamefont {Yong-Zhong}\ \bibnamefont {Qian}},\
  }\bibfield  {title} {\enquote {\bibinfo {title} {{Coherent Development of
  Neutrino Flavor in the Supernova Environment}},}\ }\href {\doibase
  10.1103/PhysRevLett.97.241101} {\bibfield  {journal} {\bibinfo  {journal}
  {Phys. Rev. Lett.}\ }\textbf {\bibinfo {volume} {97}},\ \bibinfo {pages}
  {241101} (\bibinfo {year} {2006}{\natexlab{c}})},\ \Eprint
  {http://arxiv.org/abs/astro-ph/0608050} {arXiv:astro-ph/0608050 [astro-ph]}
  \BibitemShut {NoStop}%
\bibitem [{\citenamefont {Fogli}\ \emph {et~al.}(2007)\citenamefont {Fogli},
  \citenamefont {Lisi}, \citenamefont {Marrone},\ and\ \citenamefont
  {Mirizzi}}]{Fogli:2007bk}%
  \BibitemOpen
  \bibfield  {author} {\bibinfo {author} {\bibfnamefont {Gianluigi~L.}\
  \bibnamefont {Fogli}}, \bibinfo {author} {\bibfnamefont {Eligio}\
  \bibnamefont {Lisi}}, \bibinfo {author} {\bibfnamefont {Antonio}\
  \bibnamefont {Marrone}}, \ and\ \bibinfo {author} {\bibfnamefont
  {Alessandro}\ \bibnamefont {Mirizzi}},\ }\bibfield  {title} {\enquote
  {\bibinfo {title} {{Collective neutrino flavor transitions in supernovae and
  the role of trajectory averaging}},}\ }\href {\doibase
  10.1088/1475-7516/2007/12/010} {\bibfield  {journal} {\bibinfo  {journal}
  {JCAP}\ }\textbf {\bibinfo {volume} {0712}},\ \bibinfo {pages} {010}
  (\bibinfo {year} {2007})},\ \Eprint {http://arxiv.org/abs/0707.1998}
  {arXiv:0707.1998 [hep-ph]} \BibitemShut {NoStop}%
\bibitem [{\citenamefont {Duan}\ \emph {et~al.}(2007)\citenamefont {Duan},
  \citenamefont {Fuller}, \citenamefont {Carlson},\ and\ \citenamefont
  {Qian}}]{Duan:2007bt}%
  \BibitemOpen
  \bibfield  {author} {\bibinfo {author} {\bibfnamefont {Huaiyu}\ \bibnamefont
  {Duan}}, \bibinfo {author} {\bibfnamefont {George~M.}\ \bibnamefont
  {Fuller}}, \bibinfo {author} {\bibfnamefont {J.}~\bibnamefont {Carlson}}, \
  and\ \bibinfo {author} {\bibfnamefont {Yong-Zhong}\ \bibnamefont {Qian}},\
  }\bibfield  {title} {\enquote {\bibinfo {title} {{Neutrino Mass Hierarchy and
  Stepwise Spectral Swapping of Supernova Neutrino Flavors}},}\ }\href
  {\doibase 10.1103/PhysRevLett.99.241802} {\bibfield  {journal} {\bibinfo
  {journal} {Phys. Rev. Lett.}\ }\textbf {\bibinfo {volume} {99}},\ \bibinfo
  {pages} {241802} (\bibinfo {year} {2007})},\ \Eprint
  {http://arxiv.org/abs/0707.0290} {arXiv:0707.0290 [astro-ph]} \BibitemShut
  {NoStop}%
\bibitem [{\citenamefont {Esteban-Pretel}\ \emph {et~al.}(2007)\citenamefont
  {Esteban-Pretel}, \citenamefont {Pastor}, \citenamefont {Tomas},
  \citenamefont {Raffelt},\ and\ \citenamefont {Sigl}}]{EstebanPretel:2007ec}%
  \BibitemOpen
  \bibfield  {author} {\bibinfo {author} {\bibfnamefont {Andreu}\ \bibnamefont
  {Esteban-Pretel}}, \bibinfo {author} {\bibfnamefont {Sergio}\ \bibnamefont
  {Pastor}}, \bibinfo {author} {\bibfnamefont {Ricard}\ \bibnamefont {Tomas}},
  \bibinfo {author} {\bibfnamefont {Georg~G.}\ \bibnamefont {Raffelt}}, \ and\
  \bibinfo {author} {\bibfnamefont {Gunter}\ \bibnamefont {Sigl}},\ }\bibfield
  {title} {\enquote {\bibinfo {title} {{Decoherence in supernova neutrino
  transformations suppressed by deleptonization}},}\ }\href {\doibase
  10.1103/PhysRevD.76.125018} {\bibfield  {journal} {\bibinfo  {journal} {Phys.
  Rev.}\ }\textbf {\bibinfo {volume} {D76}},\ \bibinfo {pages} {125018}
  (\bibinfo {year} {2007})},\ \Eprint {http://arxiv.org/abs/0706.2498}
  {arXiv:0706.2498 [astro-ph]} \BibitemShut {NoStop}%
\bibitem [{\citenamefont {Dasgupta}\ \emph {et~al.}(2009)\citenamefont
  {Dasgupta}, \citenamefont {Dighe}, \citenamefont {Raffelt},\ and\
  \citenamefont {Smirnov}}]{Dasgupta:2009mg}%
  \BibitemOpen
  \bibfield  {author} {\bibinfo {author} {\bibfnamefont {Basudeb}\ \bibnamefont
  {Dasgupta}}, \bibinfo {author} {\bibfnamefont {Amol}\ \bibnamefont {Dighe}},
  \bibinfo {author} {\bibfnamefont {Georg~G.}\ \bibnamefont {Raffelt}}, \ and\
  \bibinfo {author} {\bibfnamefont {Alexei~{\relax Yu}.}\ \bibnamefont
  {Smirnov}},\ }\bibfield  {title} {\enquote {\bibinfo {title} {{Multiple
  Spectral Splits of Supernova Neutrinos}},}\ }\href {\doibase
  10.1103/PhysRevLett.103.051105} {\bibfield  {journal} {\bibinfo  {journal}
  {Phys. Rev. Lett.}\ }\textbf {\bibinfo {volume} {103}},\ \bibinfo {pages}
  {051105} (\bibinfo {year} {2009})},\ \Eprint {http://arxiv.org/abs/0904.3542}
  {arXiv:0904.3542 [hep-ph]} \BibitemShut {NoStop}%
\bibitem [{\citenamefont {Friedland}(2010)}]{Friedland:2010sc}%
  \BibitemOpen
  \bibfield  {author} {\bibinfo {author} {\bibfnamefont {Alexander}\
  \bibnamefont {Friedland}},\ }\bibfield  {title} {\enquote {\bibinfo {title}
  {{Self-refraction of supernova neutrinos: mixed spectra and three-flavor
  instabilities}},}\ }\href {\doibase 10.1103/PhysRevLett.104.191102}
  {\bibfield  {journal} {\bibinfo  {journal} {Phys. Rev. Lett.}\ }\textbf
  {\bibinfo {volume} {104}},\ \bibinfo {pages} {191102} (\bibinfo {year}
  {2010})},\ \Eprint {http://arxiv.org/abs/1001.0996} {arXiv:1001.0996
  [hep-ph]} \BibitemShut {NoStop}%
\bibitem [{\citenamefont {Duan}\ \emph {et~al.}(2010)\citenamefont {Duan},
  \citenamefont {Fuller},\ and\ \citenamefont {Qian}}]{Duan:2010bg}%
  \BibitemOpen
  \bibfield  {author} {\bibinfo {author} {\bibfnamefont {Huaiyu}\ \bibnamefont
  {Duan}}, \bibinfo {author} {\bibfnamefont {George~M.}\ \bibnamefont
  {Fuller}}, \ and\ \bibinfo {author} {\bibfnamefont {Yong-Zhong}\ \bibnamefont
  {Qian}},\ }\bibfield  {title} {\enquote {\bibinfo {title} {{Collective
  Neutrino Oscillations}},}\ }\href {\doibase
  10.1146/annurev.nucl.012809.104524} {\bibfield  {journal} {\bibinfo
  {journal} {Ann. Rev. Nucl. Part. Sci.}\ }\textbf {\bibinfo {volume} {60}},\
  \bibinfo {pages} {569--594} (\bibinfo {year} {2010})},\ \Eprint
  {http://arxiv.org/abs/1001.2799} {arXiv:1001.2799 [hep-ph]} \BibitemShut
  {NoStop}%
\bibitem [{\citenamefont {Raffelt}\ \emph {et~al.}(2013)\citenamefont
  {Raffelt}, \citenamefont {Sarikas},\ and\ \citenamefont
  {de~Sousa~Seixas}}]{Raffelt:2013rqa}%
  \BibitemOpen
  \bibfield  {author} {\bibinfo {author} {\bibfnamefont {Georg}\ \bibnamefont
  {Raffelt}}, \bibinfo {author} {\bibfnamefont {Srdjan}\ \bibnamefont
  {Sarikas}}, \ and\ \bibinfo {author} {\bibfnamefont {David}\ \bibnamefont
  {de~Sousa~Seixas}},\ }\bibfield  {title} {\enquote {\bibinfo {title} {{Axial
  Symmetry Breaking in Self-Induced Flavor Conversion of Supernova Neutrino
  Fluxes}},}\ }\href {\doibase 10.1103/PhysRevLett.113.239903,
  10.1103/PhysRevLett.111.091101} {\bibfield  {journal} {\bibinfo  {journal}
  {Phys. Rev. Lett.}\ }\textbf {\bibinfo {volume} {111}},\ \bibinfo {pages}
  {091101} (\bibinfo {year} {2013})},\ \bibinfo {note} {[Erratum: Phys. Rev.
  Lett.113,no.23,239903(2014)]},\ \Eprint {http://arxiv.org/abs/1305.7140}
  {arXiv:1305.7140 [hep-ph]} \BibitemShut {NoStop}%
\bibitem [{\citenamefont {Duan}\ and\ \citenamefont
  {Shalgar}(2015)}]{Duan:2014gfa}%
  \BibitemOpen
  \bibfield  {author} {\bibinfo {author} {\bibfnamefont {Huaiyu}\ \bibnamefont
  {Duan}}\ and\ \bibinfo {author} {\bibfnamefont {Shashank}\ \bibnamefont
  {Shalgar}},\ }\bibfield  {title} {\enquote {\bibinfo {title} {{Flavor
  instabilities in the neutrino line model}},}\ }\href {\doibase
  10.1016/j.physletb.2015.05.057} {\bibfield  {journal} {\bibinfo  {journal}
  {Phys. Lett.}\ }\textbf {\bibinfo {volume} {B747}},\ \bibinfo {pages}
  {139--143} (\bibinfo {year} {2015})},\ \Eprint
  {http://arxiv.org/abs/1412.7097} {arXiv:1412.7097 [hep-ph]} \BibitemShut
  {NoStop}%
\bibitem [{\citenamefont {Chakraborty}\ \emph {et~al.}(2016)\citenamefont
  {Chakraborty}, \citenamefont {Hansen}, \citenamefont {Izaguirre},\ and\
  \citenamefont {Raffelt}}]{Chakraborty:2015tfa}%
  \BibitemOpen
  \bibfield  {author} {\bibinfo {author} {\bibfnamefont {Sovan}\ \bibnamefont
  {Chakraborty}}, \bibinfo {author} {\bibfnamefont {Rasmus~Sloth}\ \bibnamefont
  {Hansen}}, \bibinfo {author} {\bibfnamefont {Ignacio}\ \bibnamefont
  {Izaguirre}}, \ and\ \bibinfo {author} {\bibfnamefont {Georg}\ \bibnamefont
  {Raffelt}},\ }\bibfield  {title} {\enquote {\bibinfo {title} {{Self-induced
  flavor conversion of supernova neutrinos on small scales}},}\ }\href
  {\doibase 10.1088/1475-7516/2016/01/028} {\bibfield  {journal} {\bibinfo
  {journal} {JCAP}\ }\textbf {\bibinfo {volume} {1601}},\ \bibinfo {pages}
  {028} (\bibinfo {year} {2016})},\ \Eprint {http://arxiv.org/abs/1507.07569}
  {arXiv:1507.07569 [hep-ph]} \BibitemShut {NoStop}%
\bibitem [{\citenamefont {Abbar}\ and\ \citenamefont
  {Duan}(2015)}]{Abbar:2015fwa}%
  \BibitemOpen
  \bibfield  {author} {\bibinfo {author} {\bibfnamefont {Sajad}\ \bibnamefont
  {Abbar}}\ and\ \bibinfo {author} {\bibfnamefont {Huaiyu}\ \bibnamefont
  {Duan}},\ }\bibfield  {title} {\enquote {\bibinfo {title} {{Neutrino flavor
  instabilities in a time-dependent supernova model}},}\ }\href {\doibase
  10.1016/j.physletb.2015.10.019} {\bibfield  {journal} {\bibinfo  {journal}
  {Phys. Lett.}\ }\textbf {\bibinfo {volume} {B751}},\ \bibinfo {pages}
  {43--47} (\bibinfo {year} {2015})},\ \Eprint
  {http://arxiv.org/abs/1509.01538} {arXiv:1509.01538 [astro-ph.HE]}
  \BibitemShut {NoStop}%
\bibitem [{\citenamefont {Mirizzi}(2013)}]{Mirizzi:2013rla}%
  \BibitemOpen
  \bibfield  {author} {\bibinfo {author} {\bibfnamefont {Alessandro}\
  \bibnamefont {Mirizzi}},\ }\bibfield  {title} {\enquote {\bibinfo {title}
  {{Multi-azimuthal-angle effects in self-induced supernova neutrino flavor
  conversions without axial symmetry}},}\ }\href {\doibase
  10.1103/PhysRevD.88.073004} {\bibfield  {journal} {\bibinfo  {journal} {Phys.
  Rev.}\ }\textbf {\bibinfo {volume} {D88}},\ \bibinfo {pages} {073004}
  (\bibinfo {year} {2013})},\ \Eprint {http://arxiv.org/abs/1308.1402}
  {arXiv:1308.1402 [hep-ph]} \BibitemShut {NoStop}%
\bibitem [{\citenamefont {Mangano}\ \emph {et~al.}(2014)\citenamefont
  {Mangano}, \citenamefont {Mirizzi},\ and\ \citenamefont
  {Saviano}}]{Mangano:2014zda}%
  \BibitemOpen
  \bibfield  {author} {\bibinfo {author} {\bibfnamefont {Gianpiero}\
  \bibnamefont {Mangano}}, \bibinfo {author} {\bibfnamefont {Alessandro}\
  \bibnamefont {Mirizzi}}, \ and\ \bibinfo {author} {\bibfnamefont {Ninetta}\
  \bibnamefont {Saviano}},\ }\bibfield  {title} {\enquote {\bibinfo {title}
  {{Damping the neutrino flavor pendulum by breaking homogeneity}},}\ }\href
  {\doibase 10.1103/PhysRevD.89.073017} {\bibfield  {journal} {\bibinfo
  {journal} {Phys. Rev.}\ }\textbf {\bibinfo {volume} {D89}},\ \bibinfo {pages}
  {073017} (\bibinfo {year} {2014})},\ \Eprint {http://arxiv.org/abs/1403.1892}
  {arXiv:1403.1892 [hep-ph]} \BibitemShut {NoStop}%
\bibitem [{\citenamefont {Mirizzi}\ \emph {et~al.}(2015)\citenamefont
  {Mirizzi}, \citenamefont {Mangano},\ and\ \citenamefont
  {Saviano}}]{Mirizzi:2015fva}%
  \BibitemOpen
  \bibfield  {author} {\bibinfo {author} {\bibfnamefont {Alessandro}\
  \bibnamefont {Mirizzi}}, \bibinfo {author} {\bibfnamefont {Gianpiero}\
  \bibnamefont {Mangano}}, \ and\ \bibinfo {author} {\bibfnamefont {Ninetta}\
  \bibnamefont {Saviano}},\ }\bibfield  {title} {\enquote {\bibinfo {title}
  {{Self-induced flavor instabilities of a dense neutrino stream in a
  two-dimensional model}},}\ }\href {\doibase 10.1103/PhysRevD.92.021702}
  {\bibfield  {journal} {\bibinfo  {journal} {Phys. Rev.}\ }\textbf {\bibinfo
  {volume} {D92}},\ \bibinfo {pages} {021702} (\bibinfo {year} {2015})},\
  \Eprint {http://arxiv.org/abs/1503.03485} {arXiv:1503.03485 [hep-ph]}
  \BibitemShut {NoStop}%
\bibitem [{\citenamefont {Mirizzi}(2015)}]{Mirizzi:2015hwa}%
  \BibitemOpen
  \bibfield  {author} {\bibinfo {author} {\bibfnamefont {Alessandro}\
  \bibnamefont {Mirizzi}},\ }\bibfield  {title} {\enquote {\bibinfo {title}
  {{Breaking the symmetries in self-induced flavor conversions of neutrino
  beams from a ring}},}\ }\href {\doibase 10.1103/PhysRevD.92.105020}
  {\bibfield  {journal} {\bibinfo  {journal} {Phys. Rev.}\ }\textbf {\bibinfo
  {volume} {D92}},\ \bibinfo {pages} {105020} (\bibinfo {year} {2015})},\
  \Eprint {http://arxiv.org/abs/1506.06805} {arXiv:1506.06805 [hep-ph]}
  \BibitemShut {NoStop}%
\bibitem [{\citenamefont {Capozzi}\ \emph {et~al.}(2016)\citenamefont
  {Capozzi}, \citenamefont {Dasgupta},\ and\ \citenamefont
  {Mirizzi}}]{Capozzi:2016oyk}%
  \BibitemOpen
  \bibfield  {author} {\bibinfo {author} {\bibfnamefont {Francesco}\
  \bibnamefont {Capozzi}}, \bibinfo {author} {\bibfnamefont {Basudeb}\
  \bibnamefont {Dasgupta}}, \ and\ \bibinfo {author} {\bibfnamefont
  {Alessandro}\ \bibnamefont {Mirizzi}},\ }\bibfield  {title} {\enquote
  {\bibinfo {title} {{Self-induced temporal instability from a neutrino
  antenna}},}\ }\href {\doibase 10.1088/1475-7516/2016/04/043} {\bibfield
  {journal} {\bibinfo  {journal} {JCAP}\ }\textbf {\bibinfo {volume} {1604}},\
  \bibinfo {pages} {043} (\bibinfo {year} {2016})},\ \Eprint
  {http://arxiv.org/abs/1603.03288} {arXiv:1603.03288 [hep-ph]} \BibitemShut
  {NoStop}%
\bibitem [{\citenamefont {Martin}\ \emph {et~al.}(2019)\citenamefont {Martin},
  \citenamefont {Abbar},\ and\ \citenamefont {Duan}}]{Martin:2019kgi}%
  \BibitemOpen
  \bibfield  {author} {\bibinfo {author} {\bibfnamefont {Joshua~D.}\
  \bibnamefont {Martin}}, \bibinfo {author} {\bibfnamefont {Sajad}\
  \bibnamefont {Abbar}}, \ and\ \bibinfo {author} {\bibfnamefont {Huaiyu}\
  \bibnamefont {Duan}},\ }\bibfield  {title} {\enquote {\bibinfo {title}
  {{Nonlinear flavor development of a two-dimensional neutrino gas}},}\ }\href
  {\doibase 10.1103/PhysRevD.100.023016} {\bibfield  {journal} {\bibinfo
  {journal} {Phys. Rev.}\ }\textbf {\bibinfo {volume} {D100}},\ \bibinfo
  {pages} {023016} (\bibinfo {year} {2019})},\ \Eprint
  {http://arxiv.org/abs/1904.08877} {arXiv:1904.08877 [hep-ph]} \BibitemShut
  {NoStop}%
\bibitem [{\citenamefont {Martin}\ \emph {et~al.}(2020)\citenamefont {Martin},
  \citenamefont {Yi},\ and\ \citenamefont {Duan}}]{Martin:2019gxb}%
  \BibitemOpen
  \bibfield  {author} {\bibinfo {author} {\bibfnamefont {Joshua~D.}\
  \bibnamefont {Martin}}, \bibinfo {author} {\bibfnamefont {Changhao}\
  \bibnamefont {Yi}}, \ and\ \bibinfo {author} {\bibfnamefont {Huaiyu}\
  \bibnamefont {Duan}},\ }\bibfield  {title} {\enquote {\bibinfo {title}
  {{Dynamic fast flavor oscillation waves in dense neutrino gases}},}\ }\href
  {\doibase 10.1016/j.physletb.2019.135088} {\bibfield  {journal} {\bibinfo
  {journal} {Phys. Lett.}\ }\textbf {\bibinfo {volume} {B800}},\ \bibinfo
  {pages} {135088} (\bibinfo {year} {2020})},\ \Eprint
  {http://arxiv.org/abs/1909.05225} {arXiv:1909.05225 [hep-ph]} \BibitemShut
  {NoStop}%
\bibitem [{\citenamefont {Sigl}\ and\ \citenamefont
  {Raffelt}(1993)}]{Sigl:1992fn}%
  \BibitemOpen
  \bibfield  {author} {\bibinfo {author} {\bibfnamefont {G.}~\bibnamefont
  {Sigl}}\ and\ \bibinfo {author} {\bibfnamefont {G.}~\bibnamefont {Raffelt}},\
  }\bibfield  {title} {\enquote {\bibinfo {title} {{General kinetic description
  of relativistic mixed neutrinos}},}\ }\href {\doibase
  10.1016/0550-3213(93)90175-O} {\bibfield  {journal} {\bibinfo  {journal}
  {Nucl. Phys.}\ }\textbf {\bibinfo {volume} {B406}},\ \bibinfo {pages}
  {423--451} (\bibinfo {year} {1993})}\BibitemShut {NoStop}%
\bibitem [{\citenamefont {Shalgar}()}]{Shalgar}%
  \BibitemOpen
  \bibfield  {author} {\bibinfo {author} {\bibfnamefont {Shashank}\
  \bibnamefont {Shalgar}},\ }\href@noop {} {\ }\bibinfo {note}
  {(unpublished)}\BibitemShut {NoStop}%
\bibitem [{\citenamefont {Press}\ \emph {et~al.}(2002)\citenamefont {Press},
  \citenamefont {Teukolsky}, \citenamefont {Vetterling},\ and\ \citenamefont
  {Flannery}}]{NR2002}%
  \BibitemOpen
  \bibfield  {author} {\bibinfo {author} {\bibfnamefont {William~H.}\
  \bibnamefont {Press}}, \bibinfo {author} {\bibfnamefont {Saul~A.}\
  \bibnamefont {Teukolsky}}, \bibinfo {author} {\bibfnamefont {William~T.}\
  \bibnamefont {Vetterling}}, \ and\ \bibinfo {author} {\bibfnamefont
  {Brian~P.}\ \bibnamefont {Flannery}},\ }\href@noop {} {\emph {\bibinfo
  {title} {Numerical Recipes in C++: The Art of Scientific Computing}}},\
  \bibinfo {edition} {2nd}\ ed.\ (\bibinfo  {publisher} {Cambridge University
  Press},\ \bibinfo {address} {Cambridge},\ \bibinfo {year} {2002})\BibitemShut
  {NoStop}%
\bibitem [{\citenamefont {Chakraborty}\ and\ \citenamefont
  {Mirizzi}(2014)}]{Mirizzi:2013wda}%
  \BibitemOpen
  \bibfield  {author} {\bibinfo {author} {\bibfnamefont {Sovan}\ \bibnamefont
  {Chakraborty}}\ and\ \bibinfo {author} {\bibfnamefont {Alessandro}\
  \bibnamefont {Mirizzi}},\ }\bibfield  {title} {\enquote {\bibinfo {title}
  {{Multi-azimuthal-angle instability for different supernova neutrino
  fluxes}},}\ }\href {\doibase 10.1103/PhysRevD.90.033004} {\bibfield
  {journal} {\bibinfo  {journal} {Phys. Rev.}\ }\textbf {\bibinfo {volume}
  {D90}},\ \bibinfo {pages} {033004} (\bibinfo {year} {2014})},\ \Eprint
  {http://arxiv.org/abs/1308.5255} {arXiv:1308.5255 [hep-ph]} \BibitemShut
  {NoStop}%
\bibitem [{\citenamefont {Raffelt}\ and\ \citenamefont
  {de~Sousa~Seixas}(2013)}]{Raffelt:2013isa}%
  \BibitemOpen
  \bibfield  {author} {\bibinfo {author} {\bibfnamefont {Georg}\ \bibnamefont
  {Raffelt}}\ and\ \bibinfo {author} {\bibfnamefont {David}\ \bibnamefont
  {de~Sousa~Seixas}},\ }\bibfield  {title} {\enquote {\bibinfo {title}
  {{Neutrino flavor pendulum in both mass hierarchies}},}\ }\href {\doibase
  10.1103/PhysRevD.88.045031} {\bibfield  {journal} {\bibinfo  {journal} {Phys.
  Rev.}\ }\textbf {\bibinfo {volume} {D88}},\ \bibinfo {pages} {045031}
  (\bibinfo {year} {2013})},\ \Eprint {http://arxiv.org/abs/1307.7625}
  {arXiv:1307.7625 [hep-ph]} \BibitemShut {NoStop}%
\bibitem [{\citenamefont {Duan}(2013)}]{Duan:2013kba}%
  \BibitemOpen
  \bibfield  {author} {\bibinfo {author} {\bibfnamefont {Huaiyu}\ \bibnamefont
  {Duan}},\ }\bibfield  {title} {\enquote {\bibinfo {title} {{Flavor
  Oscillation Modes In Dense Neutrino Media}},}\ }\href {\doibase
  10.1103/PhysRevD.88.125008} {\bibfield  {journal} {\bibinfo  {journal} {Phys.
  Rev.}\ }\textbf {\bibinfo {volume} {D88}},\ \bibinfo {pages} {125008}
  (\bibinfo {year} {2013})},\ \Eprint {http://arxiv.org/abs/1309.7377}
  {arXiv:1309.7377 [hep-ph]} \BibitemShut {NoStop}%
\bibitem [{\citenamefont {Duan}\ and\ \citenamefont
  {Friedland}(2011)}]{Duan:2010bf}%
  \BibitemOpen
  \bibfield  {author} {\bibinfo {author} {\bibfnamefont {Huaiyu}\ \bibnamefont
  {Duan}}\ and\ \bibinfo {author} {\bibfnamefont {Alexander}\ \bibnamefont
  {Friedland}},\ }\bibfield  {title} {\enquote {\bibinfo {title} {{Self-induced
  suppression of collective neutrino oscillations in a supernova}},}\ }\href
  {\doibase 10.1103/PhysRevLett.106.091101} {\bibfield  {journal} {\bibinfo
  {journal} {Phys. Rev. Lett.}\ }\textbf {\bibinfo {volume} {106}},\ \bibinfo
  {pages} {091101} (\bibinfo {year} {2011})},\ \Eprint
  {http://arxiv.org/abs/1006.2359} {arXiv:1006.2359 [hep-ph]} \BibitemShut
  {NoStop}%
\bibitem [{\citenamefont {Duan}\ \emph {et~al.}(2011)\citenamefont {Duan},
  \citenamefont {Friedland}, \citenamefont {McLaughlin},\ and\ \citenamefont
  {Surman}}]{Duan:2010af}%
  \BibitemOpen
  \bibfield  {author} {\bibinfo {author} {\bibfnamefont {Huaiyu}\ \bibnamefont
  {Duan}}, \bibinfo {author} {\bibfnamefont {Alexander}\ \bibnamefont
  {Friedland}}, \bibinfo {author} {\bibfnamefont {GailC.}\ \bibnamefont
  {McLaughlin}}, \ and\ \bibinfo {author} {\bibfnamefont {Rebecca}\
  \bibnamefont {Surman}},\ }\bibfield  {title} {\enquote {\bibinfo {title}
  {{The influence of collective neutrino oscillations on a supernova
  r-process}},}\ }\href {\doibase 10.1088/0954-3899/38/3/035201} {\bibfield
  {journal} {\bibinfo  {journal} {J. Phys.}\ }\textbf {\bibinfo {volume}
  {G38}},\ \bibinfo {pages} {035201} (\bibinfo {year} {2011})},\ \Eprint
  {http://arxiv.org/abs/1012.0532} {arXiv:1012.0532 [astro-ph.SR]} \BibitemShut
  {NoStop}%
\bibitem [{\citenamefont {Wu}\ \emph {et~al.}(2015)\citenamefont {Wu},
  \citenamefont {Qian}, \citenamefont {Martinez-Pinedo}, \citenamefont
  {Fischer},\ and\ \citenamefont {Huther}}]{Wu:2014kaa}%
  \BibitemOpen
  \bibfield  {author} {\bibinfo {author} {\bibfnamefont {Meng-Ru}\ \bibnamefont
  {Wu}}, \bibinfo {author} {\bibfnamefont {Yong-Zhong}\ \bibnamefont {Qian}},
  \bibinfo {author} {\bibfnamefont {Gabriel}\ \bibnamefont {Martinez-Pinedo}},
  \bibinfo {author} {\bibfnamefont {Tobias}\ \bibnamefont {Fischer}}, \ and\
  \bibinfo {author} {\bibfnamefont {Lutz}\ \bibnamefont {Huther}},\ }\bibfield
  {title} {\enquote {\bibinfo {title} {{Effects of neutrino oscillations on
  nucleosynthesis and neutrino signals for an 18 M supernova model}},}\ }\href
  {\doibase 10.1103/PhysRevD.91.065016} {\bibfield  {journal} {\bibinfo
  {journal} {Phys. Rev.}\ }\textbf {\bibinfo {volume} {D91}},\ \bibinfo {pages}
  {065016} (\bibinfo {year} {2015})},\ \Eprint {http://arxiv.org/abs/1412.8587}
  {arXiv:1412.8587 [astro-ph.HE]} \BibitemShut {NoStop}%
\bibitem [{\citenamefont {Mirizzi}\ \emph {et~al.}(2016)\citenamefont
  {Mirizzi}, \citenamefont {Tamborra}, \citenamefont {Janka}, \citenamefont
  {Saviano}, \citenamefont {Scholberg}, \citenamefont {Bollig}, \citenamefont
  {Hudepohl},\ and\ \citenamefont {Chakraborty}}]{Mirizzi:2015eza}%
  \BibitemOpen
  \bibfield  {author} {\bibinfo {author} {\bibfnamefont {Alessandro}\
  \bibnamefont {Mirizzi}}, \bibinfo {author} {\bibfnamefont {Irene}\
  \bibnamefont {Tamborra}}, \bibinfo {author} {\bibfnamefont {Hans-Thomas}\
  \bibnamefont {Janka}}, \bibinfo {author} {\bibfnamefont {Ninetta}\
  \bibnamefont {Saviano}}, \bibinfo {author} {\bibfnamefont {Kate}\
  \bibnamefont {Scholberg}}, \bibinfo {author} {\bibfnamefont {Robert}\
  \bibnamefont {Bollig}}, \bibinfo {author} {\bibfnamefont {Lorenz}\
  \bibnamefont {Hudepohl}}, \ and\ \bibinfo {author} {\bibfnamefont {Sovan}\
  \bibnamefont {Chakraborty}},\ }\bibfield  {title} {\enquote {\bibinfo {title}
  {{Supernova Neutrinos: Production, Oscillations and Detection}},}\ }\href
  {\doibase 10.1393/ncr/i2016-10120-8} {\bibfield  {journal} {\bibinfo
  {journal} {Riv. Nuovo Cim.}\ }\textbf {\bibinfo {volume} {39}},\ \bibinfo
  {pages} {1--112} (\bibinfo {year} {2016})},\ \Eprint
  {http://arxiv.org/abs/1508.00785} {arXiv:1508.00785 [astro-ph.HE]}
  \BibitemShut {NoStop}%
\bibitem [{\citenamefont {Raffelt}\ and\ \citenamefont
  {Sigl}(2007)}]{Raffelt:2007yz}%
  \BibitemOpen
  \bibfield  {author} {\bibinfo {author} {\bibfnamefont {G.~G.}\ \bibnamefont
  {Raffelt}}\ and\ \bibinfo {author} {\bibfnamefont {G.}~\bibnamefont {Sigl}},\
  }\bibfield  {title} {\enquote {\bibinfo {title} {{Self-induced decoherence in
  dense neutrino gases}},}\ }\href {\doibase 10.1103/PhysRevD.75.083002}
  {\bibfield  {journal} {\bibinfo  {journal} {Phys. Rev.}\ }\textbf {\bibinfo
  {volume} {D75}},\ \bibinfo {pages} {083002} (\bibinfo {year} {2007})},\
  \Eprint {http://arxiv.org/abs/hep-ph/0701182} {arXiv:hep-ph/0701182 [hep-ph]}
  \BibitemShut {NoStop}%
\end{thebibliography}%

\end{document}